\documentclass[12pt,a4paper, captions=figureheading]{scrartcl}
\usepackage[utf8]{inputenc}
\usepackage[T1]{fontenc}
\usepackage{lmodern}
 \usepackage{ragged2e}
\usepackage{pdfpages}
\setkomafont{disposition}{\normalcolor\bfseries\rmfamily}

\usepackage[affil-it]{authblk}
\usepackage{amsmath, amsfonts, amssymb, bbm, dsfont}
\usepackage{graphicx, color, enumitem, soul, verbatim}
\usepackage{appendix}
\usepackage[colorlinks=true, allcolors=blue]{hyperref}
\usepackage{setspace, float, placeins, adjustbox}
\usepackage{pdflscape}
\usepackage{gensymb}
\usepackage{subcaption}
\usepackage{tocloft}
\makeatletter
\renewcommand{\numberline}[1]{{\@cftbsnum #1\@cftasnum~}\@cftasnumb}
\makeatother
\setkomafont{caption}{\small}
\setkomafont{captionlabel}{\usekomafont{caption}}

\usepackage[style=authoryear, citetracker=true, maxcitenames=2, hyperref=true, url=true, isbn=false, doi=true, natbib=true, bibstyle=authoryear, maxbibnames=30, sorting=nyt,sortcites, uniquename=full, uniquelist=false]{biblatex}
\usepackage{csquotes}
\newcommand{\black}[1]{\textcolor{black}{#1}}

\setlist{nosep}

\title{
Knowledge for a warmer world:\\A patent analysis of climate change adaptation technologies
}
\author{Kerstin H\"otte$^{1,2,3}$\footnote{
Corresponding author: kerstin.hotte@oxfordmartin.ox.ac.uk\\ORCiD: KH: 0000-0002-8633-4225; SJ: 0000-0001-9582-8289
}, Su Jung Jee$^{2,4}$
\\
\footnotesize{$^{1}$ Oxford Martin Programme on Technological and Economic Change, University of Oxford, UK}\\
\footnotesize{$^{2}$ Institute for New Economic Thinking at the Oxford Martin School, University of Oxford, UK}\\
\footnotesize{$^{3}$ Faculty of Business Administration and Economics, University of Bielefeld, DE}\\
\footnotesize{$^{4}$ School of Management, Faculty of Management, Law \& Social Sciences, University of Bradford, UK}\\
}

\vspace{-0.75cm}
\date{\today}

\begin{document}
\vspace{-1.5cm}
\maketitle
\vspace{-0.75cm}
\begin{abstract}
\noindent 
\black{Technologies can help strengthen the resilience of our economy against existential climate-risks. 
We investigate climate change adaptation technologies (CCATs) in US patents to understand (1) historical patterns and drivers 
of innovation; (2) scientific and technological requirements to develop and use CCATs; and (3) CCATs' potential technological synergies with mitigation.  
First, in contrast to mitigation, innovation in CCATs only slowly takes off, indicating a relatively low awareness of investors for solutions to cope with climate risks. Historical trends in environmental regulation, energy prices, and public support can be associated with patenting in CCATs. 
Second, CCATs form two main clusters: science-intensive ones in agriculture, health, and monitoring technologies; and engineering-intensive ones in coastal, water, and infrastructure technologies. Analyses of technology-specific scientific and technological knowledge bases inform directions for how to facilitate advancement, transfer and use of CCATs. Lastly, CCATs show strong technological complementarities with mitigation as more than 25\% of CCATs bear mitigation benefits. While not judging about the complementarity of mitigation and adaptation in general, our results suggest how policymakers can harness these technological synergies to achieve both goals simultaneously.} 
\end{abstract}

\vspace{0.5cm}



\newpage
\section*{Highlights}


\begin{itemize}
    \item \black{Patenting in adaptation stagnates indicating a low awareness of investors for climate risks.} 
    \item \black{Adaptation technologies rely more strongly on public support than others.} 
   \item \black{Analysis of science base informs how to leverage regional absorptive capacity to transfer adaptation technologies.} 
    \item \black{Analysis of technological base informs how to stimulate the advancement of adaptation technologies.} 
    \item \black{Complementarities between adaptation and mitigation show scope to maximize synergies.} 

\end{itemize}

\newpage
\tableofcontents
\newpage
\section{Introduction}
Climate change is an existential threat to human livelihoods \citep{bellprat2019towards, ornes2018core}. Recent extreme weather events have shown that significant adaptation is needed to help communities, cities, and economic activity adjust to new climatic conditions \citep{ipcc2018special}. 
\color{black}Technological innovation plays an important role in addressing this challenge \citep{ferreira2020technology, dechezlepretre2020invention}: climate-smart agriculture helps to adapt to droughts, floods, and increasing threats of pest infestation \citep{Kuhl2020., adenle2015global}; new types of hazard defense and weather prediction tools help protect infrastructure and human lives from storms, floods, and heatwaves \citep{unfcc2006technologies}; water conservation and catchment technologies help address water scarcity \citep{conway2015invention}; vaccines, new drugs, and preventive public health inventions strengthen people's resistance against infectious diseases and heatwave-induced risks that become more prevalent under climate change \citep{guo2018quantifying, caminade2019impact}. \color{black}
Alongside nature-based solutions and behavioral changes, adaptation technologies are needed to cope with current and future climate risks \citep{unfcc2006technologies,unep2021adaptation}.

Governments under the Paris Agreement committed to strengthening their adaptation capacities, including technological solutions, but progress to achieve this goal has been rarely evaluated at a comprehensive level \citep{berrang2019tracking, lesnikowski2017does}. 
\black{Measuring progress is important as it helps identify adaptation gaps, enables impact assessments of adaptation strategies, and mutual learning when decision makers share their experience with undertaken adaptation efforts.
Here, we systematically take stock of existing technologies for adaptation using patent data addressing three questions: \emph{(1) 
To what extent have these technologies been developed and which were the drivers of innovation? 
(2) How can governments support the development and adoption of these technologies? (3) How do technologies for adaptation interact with climate change mitigation?} }

Existing studies on climate change adaptation technologies (CCATs) often focused on specific regions, technologies, or climate risks. 
To date, systematic analyses of innovation in CCATs have been limited \citep{popp2019environmental, dechezlepretre2020invention}, not least because, until 2018, there was no classification of CCATs in patent databases. The most closely related study was made by \citet{dechezlepretre2020invention} who investigated the diffusion of CCATs using patent data.

Leveraging the recent Cooperative Patent Classification (CPC) of `climate change adaptation technologies' \citep{angelucci2018supporting}, we investigated how innovation in various adaptation technologies has changed over time.  
We analyzed the composition of the scientific and technological knowledge bases of CCATs \black{to show which scientific and technological capabilities are needed to advance, adopt, and utilize CCATs.} We further identified technological complementarities between adaptation and mitigation \black{showing in which areas both targets can be achieved at the same time.} 
To the best of our knowledge, this is the first systematic analysis of the current state of technological knowledge for adaptation. 
From our analysis of CCATs, we have documented five key insights: 

\begin{enumerate}
  \item \black{Despite increased awareness of climate change, patenting in most adaptation technologies has not experienced a surge in the past two decades, much unlike patenting in climate change mitigation which has been increasing significantly.}
  \item Adaptation technologies form two clusters: those that are science-intensive (health-related adaptation, agriculture, and indirectly enabling technologies for weather forecasting and natural resource assessment) and those that are engineering-based (adaptation in coastal, infrastructure, and water supply). \black{The qualitative details of the knowledge base reveal scientific and technological requirements needed to develop, adopt, and utilize these technologies and inform policy makers how to facilitate advancement and transfer of adaptation technologies.}
  \item Invention in various CCATs greatly differ by magnitude: Adaptation related to human health has the highest number of patented inventions ($>$16k patents), followed by agriculture (8k). Coastal adaptation has the lowest number of patented inventions ($<$0.9k). 
  \item Since mid-2000s, more than 40\% of adaptation patents have been reliant on government support, which is about 10\% higher than average \citep{fleming2019government}. 
  For most mitigation technologies, the reliance on government support is much lower during the same period, except for nascent mitigation technologies such as carbon capture and storage (CCS).\footnote{The technology class CCS also includes the capture and storage of non-carbon greenhouse gases such as $SO_2$ or $SF_6$. We use the term CCS to simplify the notation.}
  \item 26\% of all adaptation technologies simultaneously help with mitigation. The highest overlap exists in infrastructure, where 70\% of adaptation patents also help reduce emissions. \black{Many of these inventions came as byproducts of innovations developed to cope with environmental regulation and high energy prices.}
\end{enumerate}


The observation that 26\% of adaptation technologies simultaneously contribute to mitigation is of high theoretical and practical importance. In many theoretical discourses, climate change adaptation and mitigation were treated as substitutes \citep{barrett2020dikes, reyer2017turn}. Our results question this perspective. We argue that well-designed policy can encourage innovations that meet the twin goals of adaptation and mitigation simultaneously.

We documented a substantial scope for technological complementarity \black{between certain adaptation and mitigation options. While not promising a universal solution for all adaptation and mitigation options, we illustrate examples of how emission-increasing \emph{maladaptation} can be avoided \citep{barnett2010maladaptation}.\footnote{Note that the definition of maladaptation goes beyond emission-increasing adaptation but includes any adaptation action with adverse side effects. In this article, we refer to subset of emission-increasing adaptation when using the term maladaptation.
} 
For example, thermal insulation in buildings achieves both: adaptation to heatwaves and emission-reduction through energy savings, while air-conditioning would be an example of maladaptation. Energy-intensive desalination to cope with water scarcity, another example of maladaptation, can be complemented with the integrated use of solar PV. Our analysis identifies additional cases, for example in agriculture, infrastructure, and clean production where public R\&D support can encourage inventions that meet adaptation and mitigation goals at the same time.}

\black{As technological development is path-dependent \citep{arthur1994increasing, ruttan1997induced}, subsequent technological development cumulatively builds on pre-existing technology and knowledge. Technology choices in the early phase of development are essential to prevent lock-in effects in adaptation options that undermine mitigation efforts or in mitigation strategies that increase vulnerability against climate change.}

The remainder of this paper is structured as follows: In the next section \ref{sec:background}, we offer an introduction to the economics of adaptation, mitigation, and technology. In section \ref{sec:methods} we describe the methodology and data. Section \ref{sec:results} outlines the results, first documenting innovation trends in adaptation (Section \ref{subsec:timeseries}), continuing with an analysis of the technological and scientific base (Section \ref{sec:science_base}), and ending with an analysis of adaptation-mitigation complementarity (Section \ref{sec:mitigation_adaptation}). Section \ref{sec:discussion} concludes.  

\section{Background}
\label{sec:background}
\subsection{{Adaptation and the role of technology}}
\black{Governments typically employ a portfolio of different actions to adapt to climate change. For instance, these portfolios can comprise of behavioral and nature-based solutions, technological adaptation of physical infrastructure, and insurance-like mechanisms that facilitate the economic recovery after the occurrence of an extreme event \citep{berrang2019tracking}.}

\black{Behavioral solutions can comprise of awareness and information campaigns that strengthen the risk-preparedness in the face of wildfires, storms, and floods; or teach the population about appropriate behavior during heatwaves \citep{van2019meta}. Nature-based solutions for adaptation either strengthen the resilience of ecological systems, such as through biodiversity protection, or leverage the provision of ecosystem services for water supply or green zones to alleviate heatwaves in urban areas \citep{seddon2020understanding, sharifi2021co}. 
Technologies for adaptation comprise both high-tech and low-tech solutions, and even non-patented technological solutions
\citep{dechezlepretre2020invention, ipcc2022adaptation, unfcc2006technologies}. 
Next to these, financial instruments and social safety nets play a crucial role, as financial and economic capabilities are essential to enable recovery after extreme weather events. These instruments consist of, for example, weather insurances in agriculture or real-estate, but also public recovery schemes. Furthermore, poverty reduction is an effective adaptation strategy, which is most prevalent in low-income countries \citep{linnerooth2015financial}.} 

\black{However, these adaptation options interact and mutually enhance their effectiveness. Behavioral risk-preparedness is easier to achieve if technologies provide reliable weather forecasts \citep{van2019meta}, and the costs of financial insurance schemes can be significantly reduced if technological adaptation strengthens the resilience of physical assets against extreme weather \citep{mills2007synergisms}.} 

\black{In this study, we focus on patented technological solutions for adaptation. 
Technological solutions are appealing when other adaptation options are prohibitively expensive or infeasible, and they bear the potential to overcome some of the limits to adaptation. For example, two-thirds of the world's cities are located on coastlines, which are vulnerable to sea level rise. Relocating assets and citizens is often infeasible or prohibitively expensive due to financial and social constraints \citep{fankhauser2016}. Many nature-based solutions like the restoration of mangroves for flood protection, agroforestry dealing with water scarcity, or green zones in cities to alleviate heatwaves, only work provided that extreme weather events are sufficiently moderate \citep{thomas2021global}. In a world with ongoing climate change as currently projected \citep{steffen2018trajectories, ipcc2018special}, societies need to prepare for weather events that exceed these limits. In these situations, technological solutions can play an important role \citep{tompkins2018documenting, ipcc2022adaptation}.} 

\subsection{{Adaptation and mitigation: Substitutes or complements?}}

\black{Although it has been said that ---given our current knowledge--- mitigation remains the cheapest and best adaptation, climate change is already happening today at a worrying pace and societies need to adapt to these unavoidable changes.}

\black{In the literature, the relationship between climate change mitigation and adaptation is an interesting controversy. Theoretical studies suggest that mitigation and adaptation efforts can be considered as strategic substitutes, as increased mitigation efforts reduce the need for future adaptation, while future adaptation may compensate for the lack of mitigation today \citep{barrett2020dikes, reyer2017turn, buob2011mitigate, van2011use}. 
Game theoretical considerations suggest that policymakers' ambitions to mitigate climate change may be undermined by the prospect of future technological solutions that neutralize the negative impact of climate change \citep{barrett2020dikes, buob2011mitigate}.} 

\black{This line of argument was believed to undermine the progress of international climate negotiations about mitigation and underpinned ethical concerns about research on climate engineering \citep{svoboda2017ethics} and adaptation \citep{reyer2017turn}. 
The controversy about adaptation-mitigation trade-offs is of very practical relevance today, acknowledging that adaptation is a necessity of both today and the future \citep{ipcc2022adaptation, barnett2010maladaptation}. 
Research has shown that short-term mitigation policies may undermine the future adaptation. For example, the production of carbon-neutral biofuels or rapid deforestation to sequester carbon may come with the cost of biodiversity losses, which may be essential to assist ecological systems in adapting to changing climatic conditions \citep{jeswani2020environmental, chisholm2010trade}. Other examples of maladaptation are emission-increasing solutions for adaptation, such as energy-intensive desalination techniques to improve water supply or air-conditioning in response to heatwaves \citep{barnett2010maladaptation}.} 

\black{However, theoretical models upon which the trade-off considerations build are difficult to calibrate for three major reasons:}

\black{(1) The models explore a trade-off between current costs of mitigation compared to future costs of adaptation. 
This valuation is highly sensitive to the appropriate choice of the discount rate which is empirically controversial \citep{gollier2014long}. 
Moreover, those making decisions about adaptation and mitigation may be disparate as adaptation benefits are mostly locally specific, and often private, while climate change mitigation contributes to a global (uncertain) public good \citep{abidoye2021economics}.} 

\black{(2) The economic impact of climate change is subject to uncertainty: Once tipping thresholds in the climate system are crossed, it may become unpredictable and an existential threat to human livelihood, which may be beyond the scope of any available and expected technological solutions \citep{lenton2019climate}.} 

\black{(3) Existing models suggest that investments made in mitigation cannot be spent on adaptation and vice versa. However, empirically mitigation and adaptation are not necessarily mutually exclusive and examples exist where adaptation efforts contribute to mitigation and vice versa \citep{sharifi2021co, spencer2017case, berry2015cross}.} 


\black{We provide evidence that the trade-off consideration may need to be reconsidered as we identify adaptation-mitigation complementarities in R\&D and show in which areas these co-benefits can be harnessed. In addition, we state that some examples of emission-increasing maladaptation can be a matter of technology choice, for example in desalination or air-conditioning \citep[see][]{barnett2010maladaptation}.}

\FloatBarrier
\section{Methods}
\label{sec:methods}

\subsection{Data sources}
\label{methods:data}
We used US patent data from the US Patent and Trademark Office (USPTO) and GooglePats compiled for an earlier project \citep{hotte2021rise, hotte2021data}. We used USPTO data since most high-value inventions are filed in the US, and US patents can be regarded as a good proxy for the global technological frontier \citep{albino2014understanding}. 
To ensure the uniqueness of inventions, we used the patent DOCDB family ID of patents downloaded from PATSTAT (Spring 2021 version) as the unit of analysis \citep{kang2016patstat, patstat_data}.\footnote{Throughout this document, we use simple DOCDB patent families as the unit of analysis, but we use `patent' for `patent family' as shorthand.} 
We supplemented the patents with CPC classifications obtained from the master classification file (April 2021 version) provided by USPTO.\footnote{\url{https://bulkdata.uspto.gov/data/patent/classification/cpc/}} 
To identify adaptation and mitigation technologies, we used the CPC Y02-tags \citep{angelucci2018supporting, su2017does}. 

We obtained 37,341 unique patents that are tagged as \emph{technologies for adaptation to climate change} as indicated by the CPC tag {Y02A}. We categorize them as patents for coastal adaptation ({Y02A1}), water supply and conservation ({Y02A2}), infrastructure resilience ({Y02A3}), agriculture ({Y02A4}), human health protection ({Y02A5}), and indirectly enabling technologies such as weather forecasting, monitoring, and water-resource assessment ({Y02A9}) (see \ref{SI:y02_definitions} for a detailed overview).
We also sourced mitigation-related patents distinguishing technologies at the 4-digit level (buildings ({Y02B}), CCS ({Y02C}), energy-saving ICT ({Y02D}), clean energy ({Y02E}), clean production ({Y02P}), clean transportation ({Y02T}), and clean waste ({Y02W})\citep[see][]{veefkind2012new, angelucci2018supporting}. 
The tagging scheme for climate change mitigation and adaptation technologies is based on the search algorithms that identify mitigation- and adaptation-related CPC symbols, IPC symbols, and keywords \citep[see][]{veefkind2012new, angelucci2018supporting}. Our analysis relies on the CPC version from April 2021.  

The USPTO regularly re-classifies patents whenever a new version of the CPC system becomes available. Hence, old and new patents are assigned to technology classes according to uniform principles \citep{lafond2019long}. This enables the identification of the technological ancestors of today's inventions. 
For example, early patented inventions in windmills are the technological ancestors of today's high-tech wind turbines \citep[cf.][]{hotte2021rise}. An adaptation-related example is health-related patents for improvements in medical compounds developed in the late nineteenth century to fight cholera. These inventions build the foundations of today's technology to fight infectious diseases. Similarly, many inventions in agriculture (e.g. ecological buffer zones or organic fertilizers), water conservation, and insulation in buildings have their origins in the nineteenth century. 

Some of the patents in our data serve multiple adaptation purposes. We double-counted these patents, arguing that knowledge is non-rival and patents that serve multiple adaptation purposes contribute equally to the knowledge base of these CPC 6-digit categories. This argument is further supported by the high variation in the value of patents: patents with a higher number of co-classifications tend to represent more valuable inventions \black{\citep{lerner1994importance, sun2020big, mendez2021measuring}}. 
In our data, 295 out of 37,341 unique patent families are multi-purpose adaptation technologies, i.e. are assigned to two or more 6-digit Y02A-classes.\footnote{Note that we applied the same double-counting rule for mitigation patents that are classified into multiple 4-digit subclasses of Y02.}

We further supplemented the data with information on the reliance of individual patents on governmental support \citep{fleming2019government, DVN/DKESRC_2019}. Patents are defined as being reliant on governmental support if at least one of the following five conditions hold: (1) The patent is directly owned by a governmental institution. (2) Governmental support is explicitly acknowledged in the patent document. (3) The patent cites a patent owned by a governmental institution or that acknowledged governmental support. (4) The patent cites research published by a governmental institution. (5) The patent cites research where governmental support is mentioned in the acknowledgments. \black{The first two conditions are related to the patents created by direct financial support of the government and the last three conditions are related to the patents being reliant on prior knowledge created by financial support of the government. Although there can be other routes of governmental support such as human- or facility-based ones, we focus on the direct financial support and prior knowledge base support, main areas of governmental support that can be captured comprehensively at the patent-level.}   

To analyze the technological and scientific knowledge base, we used data on (1) citations from patents to patents from \citet{pichler2020technological} and (2) citations from patents to science provided by \citet{marx2020v30}. 

\subsection{The scientific base}
\label{methods:science}
To describe the scientific base of adaptation, we used data on citations from patents to science. Citations in a patent can be made in the text body or at the front page of a patent, and can be added by the applicant or patent examiner. We included all types of citation into our analyses. \citet{marx2019reliance} extracted the citation links using a sequential procedure based on text recognition and matched the data with the scientific database Microsoft Academic Graph (MAG) \citep{sinha2015overview}. The matching procedure is probabilistic. \citet{marx2019reliance} tagged citation links with a so-called confidence score, which indicates the precision and recall rate of the matching \citep[more detail available in][]{marx2019reliance_working_paper, marx2020reliance}. We only included patents with a confidence score $>4$ which is associated with a precision rate of more than 99\%. 
The citation links are complemented with meta-information on the scientific paper that is cited, e.g. title, DOI (if available), outlet, publication year, and scientific field of research. 

We analyzed the scientific base in two ways: (1) We computed time series of the share of the number of citations to papers to the number of total citations to patents and papers. The data are aggregated into 5-year bins gathering all patents classified as certain adaptation technology that were granted during the considered period. (2) We provided a qualitative description of the science base. Every paper is tagged by the Web-of-Science (WoS) category into which the article is classified. The assignment of WoS categories to papers is made on the paper level \citep[further explanations are provided in][]{hotte2021rise}. We used this information to show, for each type of adaptation technology, the six most often cited WoS fields as a share in all scientific citations during the different time periods. 


\subsection{Adaptation-mitigation complementarities}
\label{methods:MA_complements}
Analyzing the technological base of patents relies on the hierarchical CPC system, which classifies patents into broad sections (A-H, Y) which are sequentially sub-divided into classes, sub-classes, groups, and sub-groups. The section `Y' is a special, cross-sectional tagging scheme that is used to identify climate-related technologies. We removed `Y10'-tags from our analysis because these tags are assigned to patents for technical reasons (for example to ensure  compatibility with other classification systems).  

Patents can be classified into multiple CPC classes. Co-classification indicates interdependence among different technologies. We used co-classification data to identify patents that are tagged as adaptation \textit{and} mitigation technologies. We call these patents `dual purpose' technologies. 

 
To better understand overlaps in the knowledge base of adaptation and mitigation technologies, we used backward citations. The cosine similarity of two technology types is computed as the normalized dot product of the vectors of backward citation shares made to 4-digit CPC classes for the technological similarity and to WoS-fields for  scientific similarity. We rely on the  methodology used in \cite{hotte2021rise} to create similarity networks. We illustrated the pairs of 6-digit mitigation and adaptation complements that show strong overlaps in their technological and scientific knowledge base. 

\FloatBarrier

\section{Results}
\label{sec:results}
\subsection{A slow start for adaptation}
\label{subsec:timeseries}
We analyzed the technological frontier in climate change adaptation by looking at patents granted by the USPTO that are tagged as \emph{technologies for adaptation to climate change}. This leads to a population of 37,341 unique patent families that are explicitly recognized as technologies that help in climate change adaptation. We also collected 408,348 unique patent families related to climate change mitigation to explore the technological relationship between mitigation and adaptation. 

Currently, there are six main categories of climate change adaptation patents: (1) coastal adaptation, (2) water supply and conservation, (3) infrastructure resilience, (4) agriculture, (5) human health protection and (6) indirect adaptation i.e. measurement technologies such as weather forecasting, monitoring invasive species, and water-resource assessments (see \cite{unfcc2006technologies} and \ref{SI:y02_definitions} for details).

\begin{figure}
{    \centering
    \caption{Patented inventions in adaptation technologies over time}
  
    \begin{subfigure}{.45\textwidth}
    \centering
    \caption{Mitigation \& adaptation counts}
    \label{subfig:ts_count_all_green}
    \includegraphics[width=\textwidth]{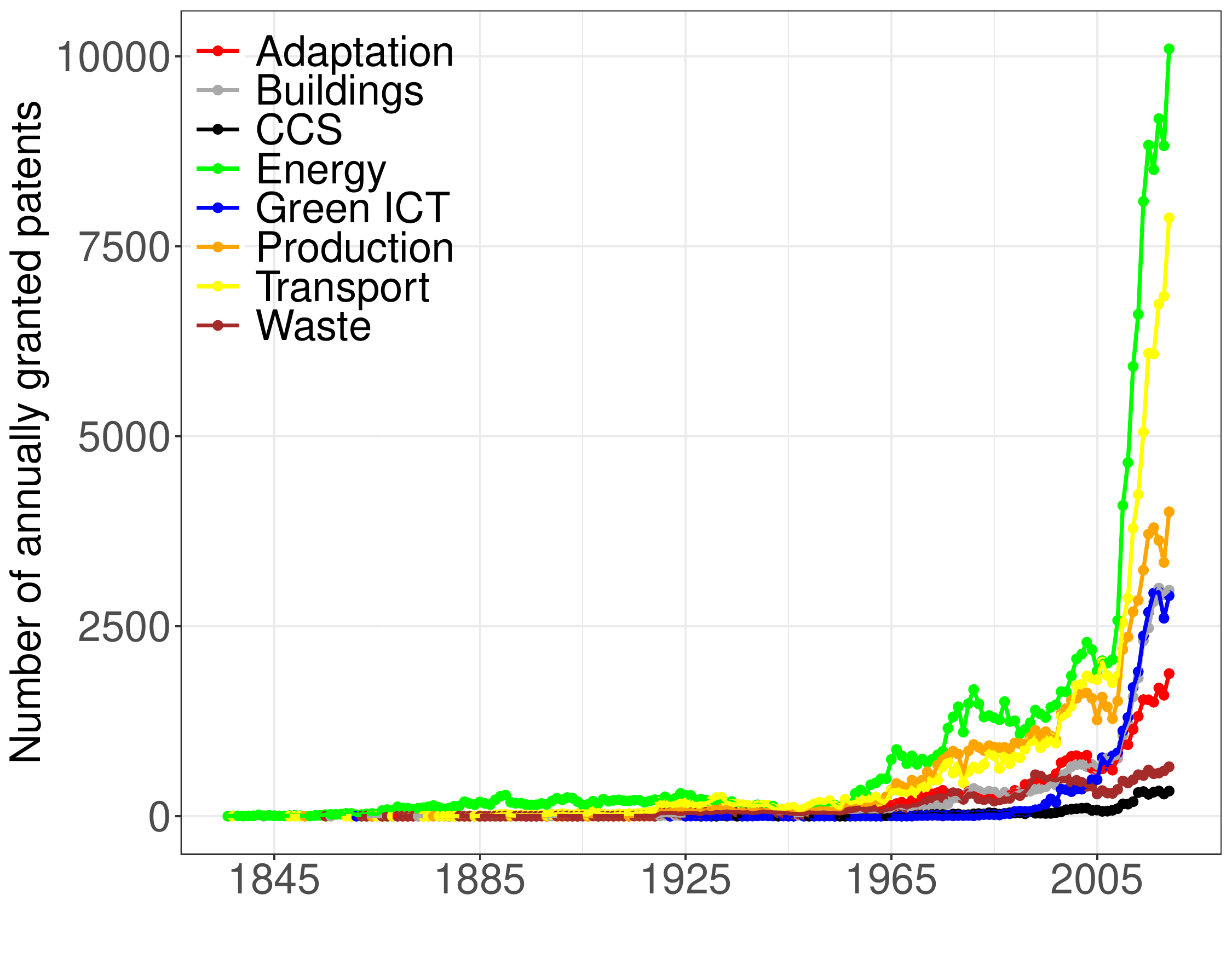}
    \end{subfigure}
    \begin{subfigure}{.45\textwidth}
    \centering
    \caption{Adaptation patents counts}
    \label{subfig:ts_count_adaptation}
    \includegraphics[width=\textwidth]{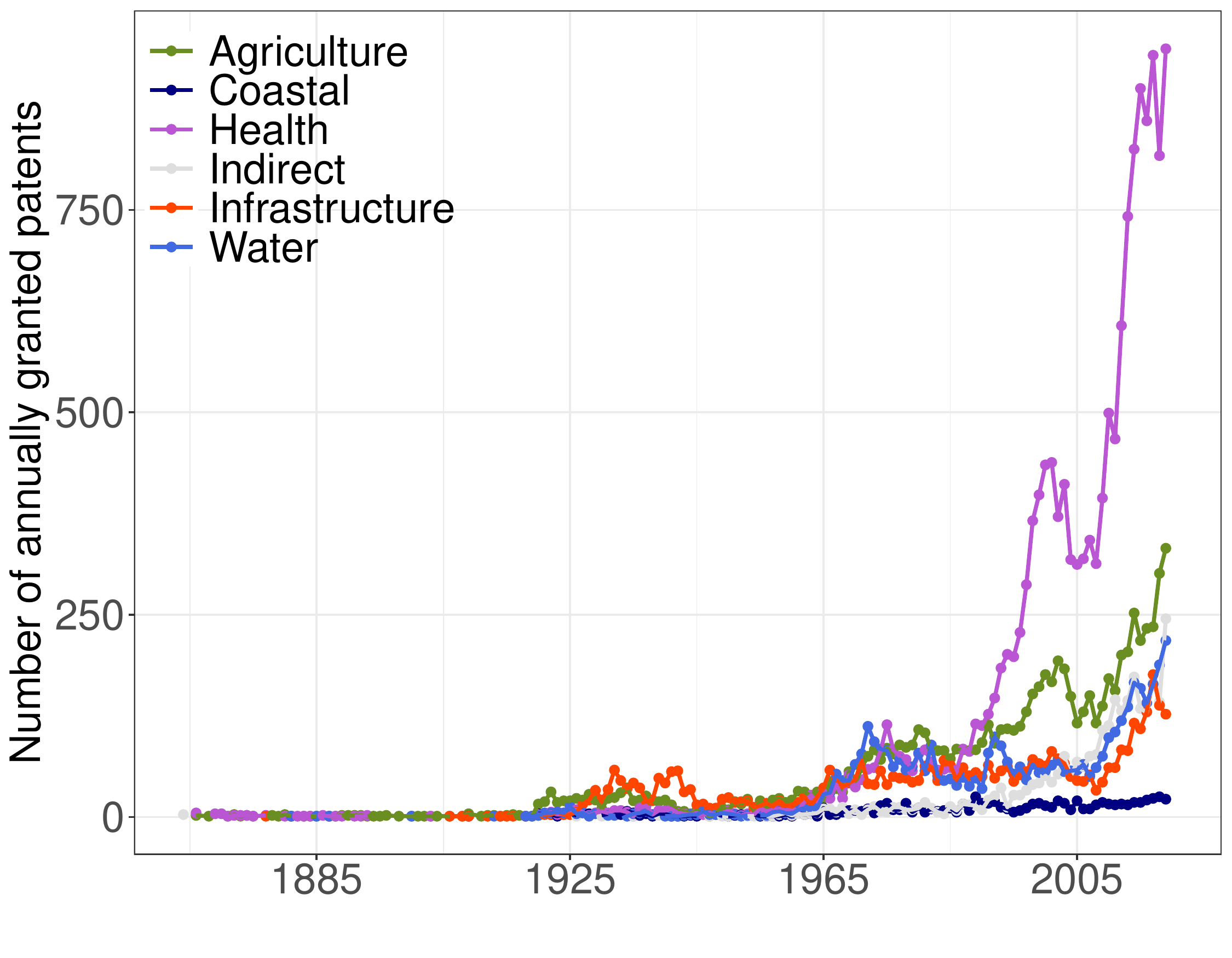}
    \end{subfigure}
    
    \centering
    \begin{subfigure}{.45\textwidth}
    \centering
    \caption{Mitigation \& adaptation shares}
    \label{subfig:ts_share_all_green}
    \includegraphics[width=\textwidth]{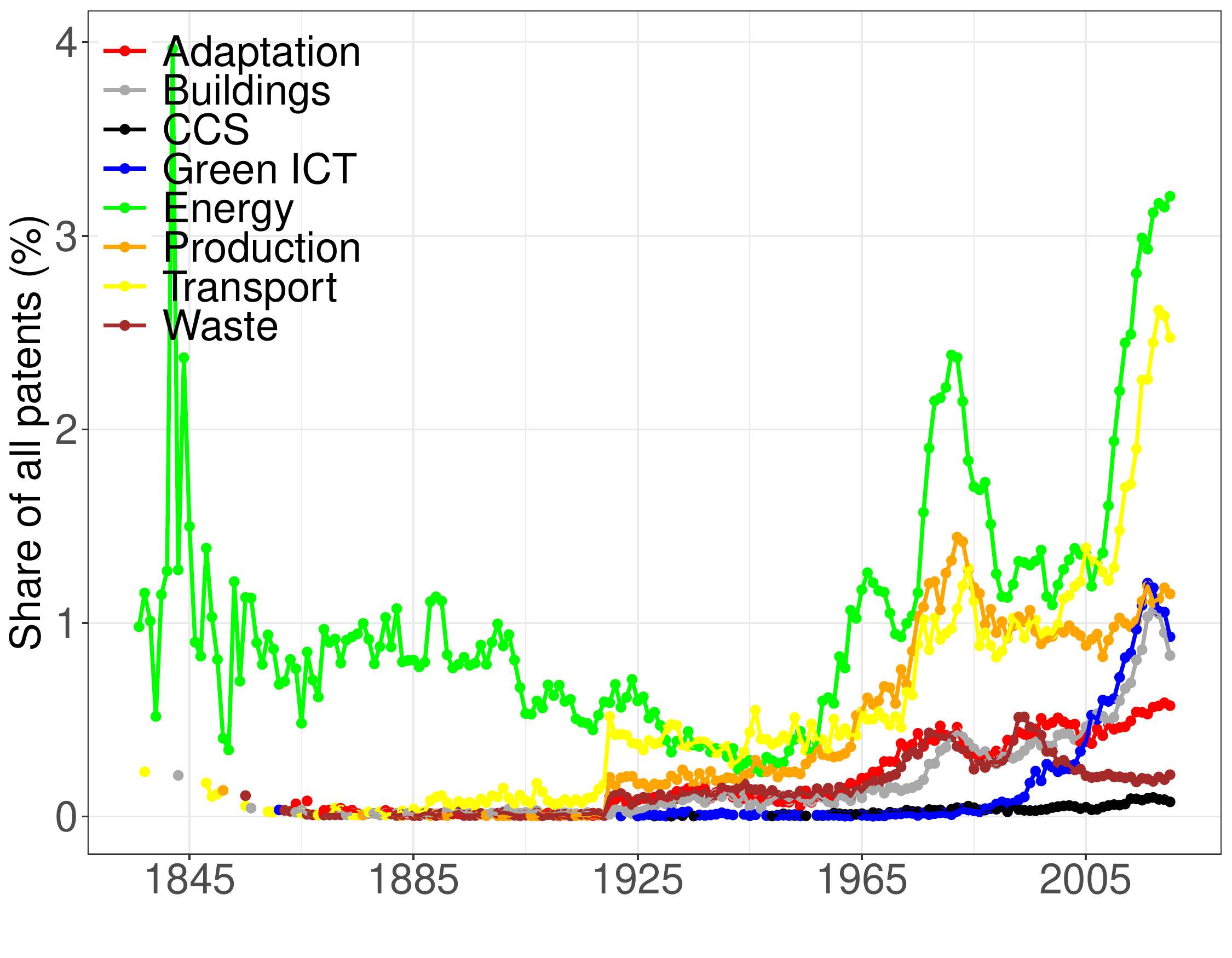}
    \end{subfigure}
    \begin{subfigure}{.45\textwidth}
    \centering
    \caption{Adaptation technology shares}
    \label{subfig:ts_share_tech1}
    \includegraphics[width=\textwidth]{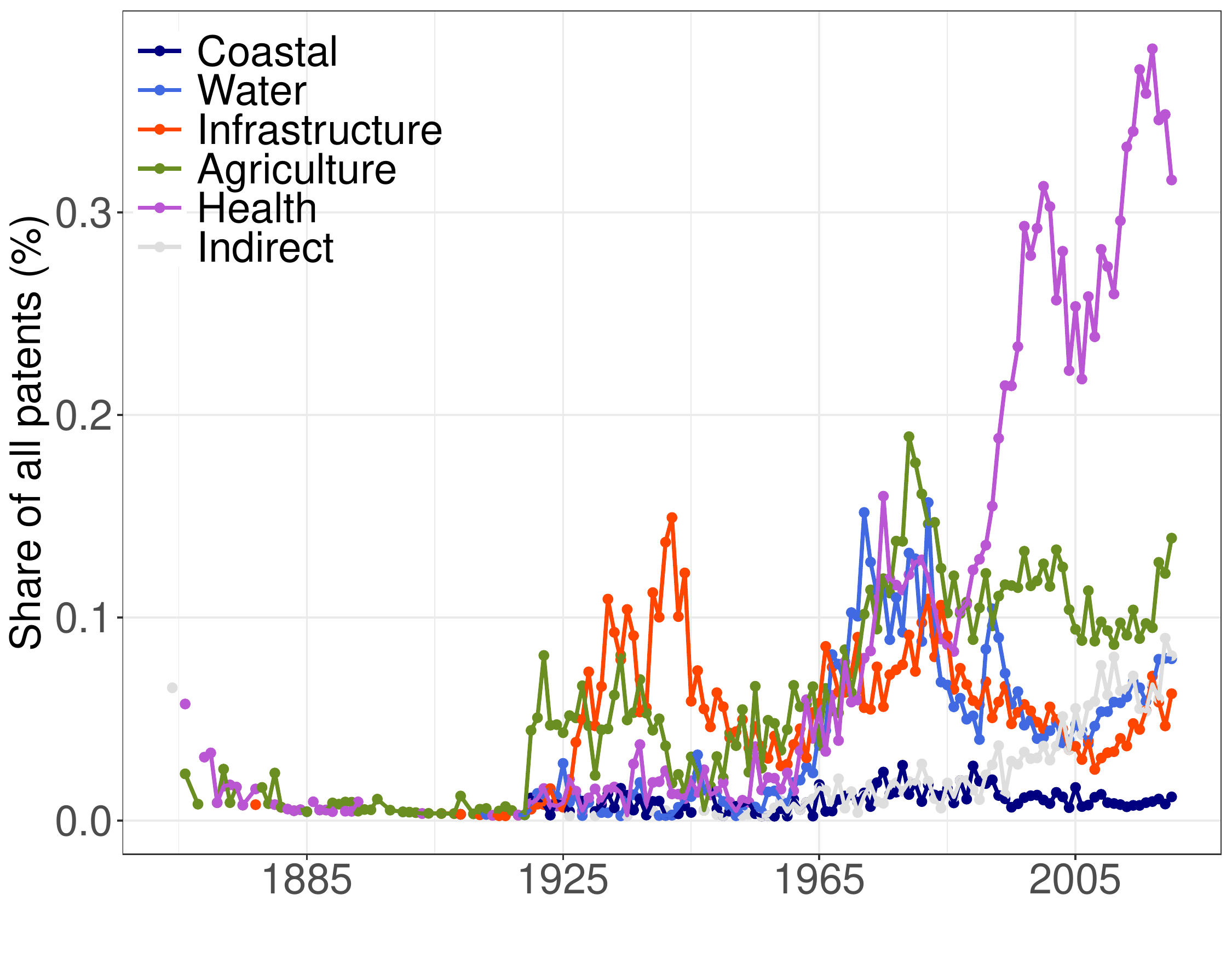}
    \end{subfigure}
    
    \centering
    \begin{subfigure}{.45\textwidth}
    \centering
    \caption{Reliance on public support}
    \label{subfig:ts_share_green_gov_supp}
    \includegraphics[width=\textwidth]{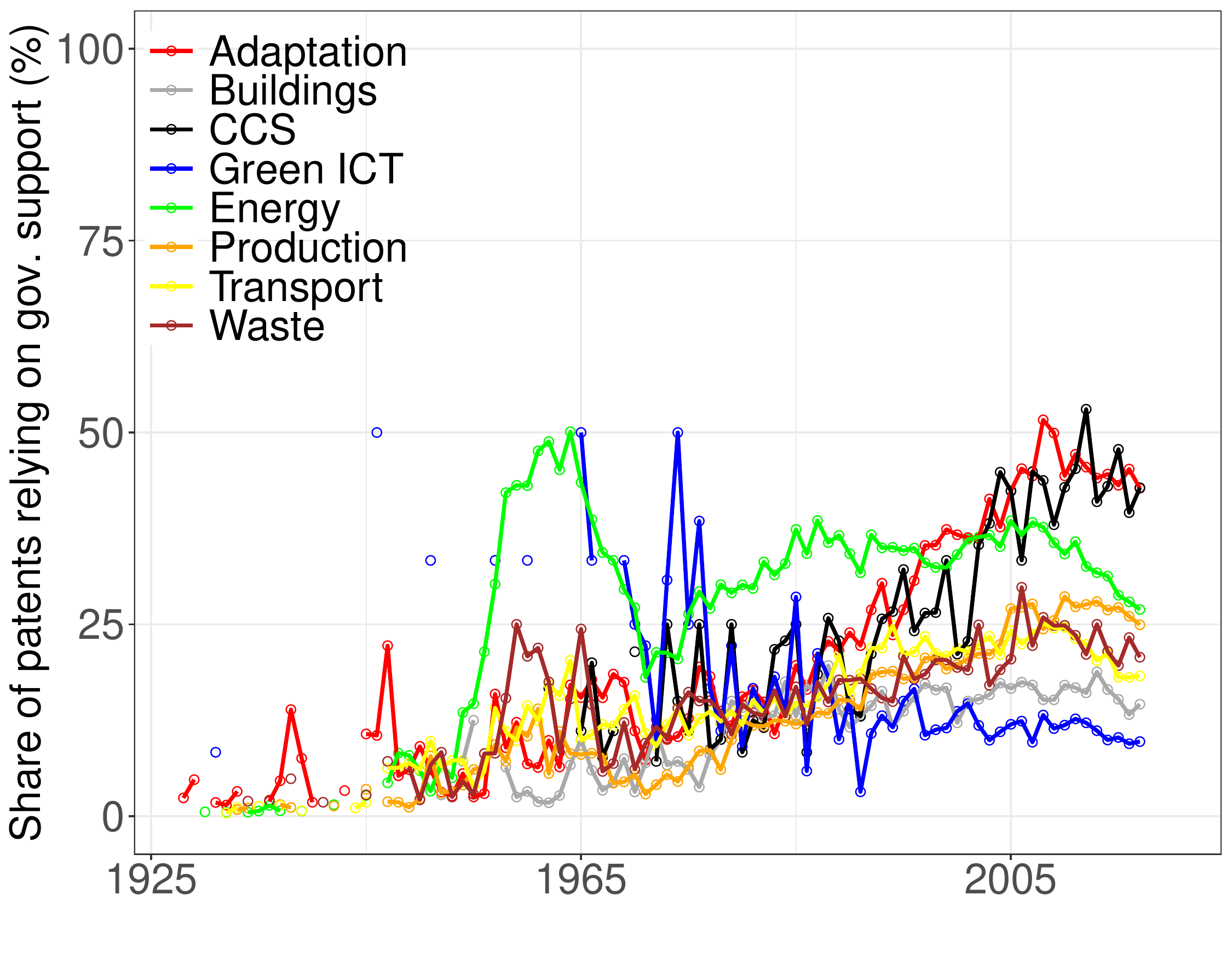}
    \end{subfigure}
    \begin{subfigure}{.45\textwidth}
    \centering
    \caption{Adaptation's reliance on public support}
    \label{subfig:ts_share_adap_gov_supp}
    \includegraphics[width=\textwidth]{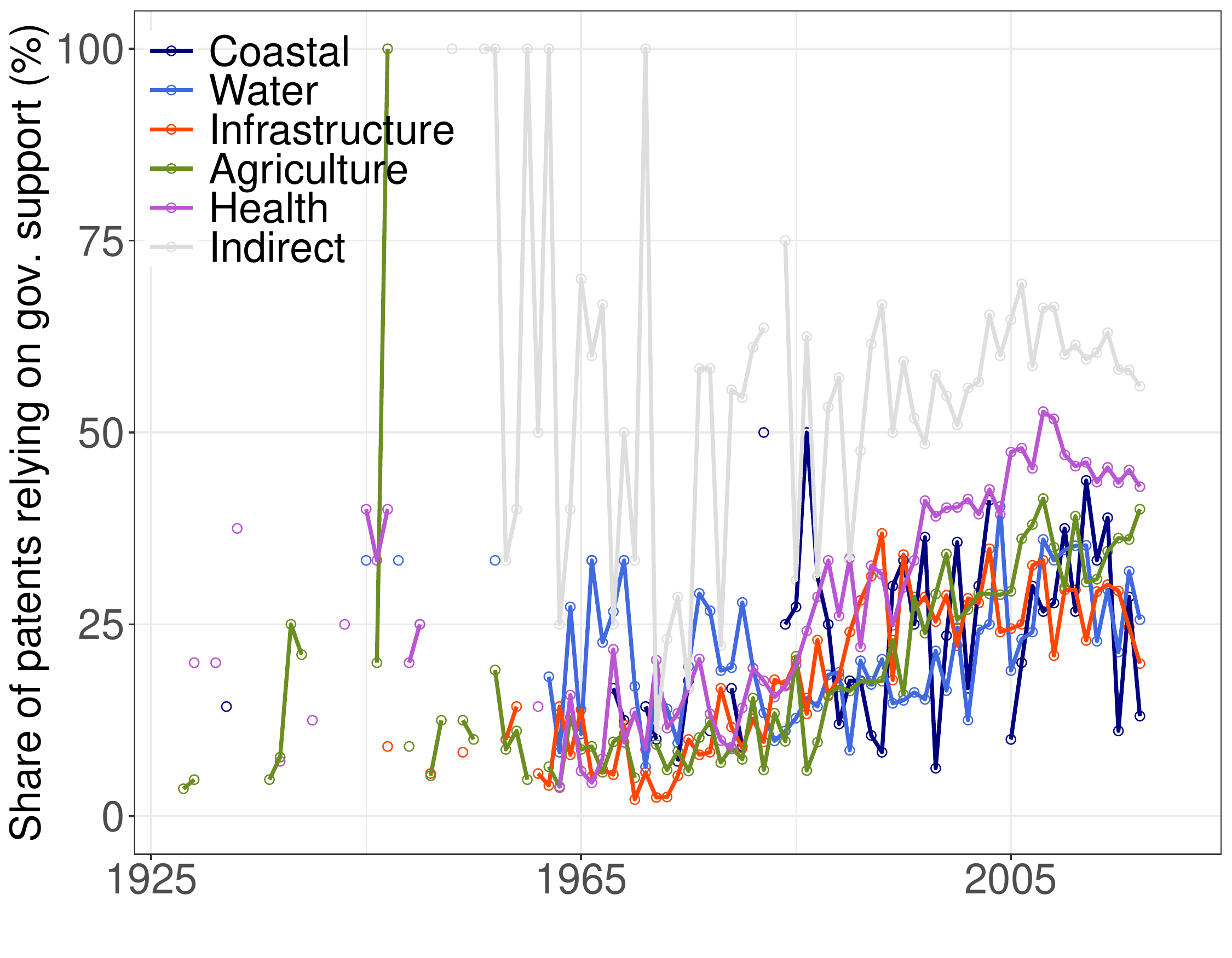}
    \end{subfigure}
    
        \label{fig:ts_patents}

}

\scriptsize
Fig. \ref{subfig:ts_count_all_green} (\ref{subfig:ts_count_adaptation}) shows the number of annually granted US patents (unique by DOCDB patent family) in 4-digit mitigation and adaptation (6-digit adaptation) technologies since 1836. 
Fig. \ref{subfig:ts_share_all_green} (\ref{subfig:ts_share_tech1}) shows the number of 4-digit mitigation and adaptation (6-digit adaptation) patents as a share of all US patents. 
Fig. \ref{subfig:ts_share_green_gov_supp} (\ref{subfig:ts_share_adap_gov_supp}) shows the share of these patents that relies on governmental support in 1935-2017 (see Sec. \ref{methods:data} for a definition). Note that the axes may differ in scale due to differences in the data by time coverage and scale. 
\end{figure}

\black{
In Fig. \ref{fig:ts_patents}, we show on the left-hand side the evolution of patents in mitigation and adaptation technologies as identified by 4-digit CPC codes. At the right-hand side, we show analogous figures for different adaptation technologies at the more disaggregate 6-digit level. 
The upper two figures \ref{subfig:ts_count_all_green} and \ref{subfig:ts_count_adaptation} show the number of annually granted patents since the mid-nineteenth century. In the mid row, we show these time series measured as a share in all annually granted USPTO patents.} 

Until the second half of the twentieth century, patenting in mitigation and adaptation ranged at very low levels, both in absolute patent counts and measured as a share. The only exception is renewable energy with a share of up to 4\% already in the late nineteenth century. This share corresponds to the level of clean energy inventions today. The historically high share is in line with previous research and historical accounts showing that patenting for energy technologies like windmills and water wheels was already very prevalent in the nineteenth century \citep{hotte2021rise}. 

Since the early twentieth century, we observe the number of annually granted patents in mitigation and adaptation to grow slowly. However, when showing these inventions as a share in all US inventions, we find the growth of green inventions to be non-monotonous. 

Patenting in climate change mitigation began growing after the 1950s. 
After then, patenting in mitigation technologies experienced several periods of acceleration, such as during the Oil Crisis of the 1970s \citep{geels2017socio, grubler2012policies}. Since the 2000s, patenting in climate change mitigation (especially in energy and transportation) increasingly took off. 

For adaptation, we find that inventions have not increased substantially over time except for the category of health-related technologies (Fig. \ref{subfig:ts_share_tech1}). Adaptation has seen only modest increases in response to the oil price crisis and in the subsequent decades. In 2020, about 0.5\% of all US patents were classified as being helpful for adaptation, while green energy and transport patents account for more than 3\% and 2.5\% and exibited a steep increase since the 2000s (Fig. \ref{subfig:ts_share_all_green}, \ref{subfig:ts_share_tech1}). 

Among the adaptation technologies, health-related adaptation dominates by the number of patents. With more than 16,300 unique patents over the full time horizon, health accounts for more than twice as many patents as agriculture (8,089 patents), which is the second largest category. Coastal adaptation is the smallest category with only 857 unique patents (Table \ref{tab:overview_adaptation_patents}). These differences may not only reflect high levels of innovative activity in health-related adaptation but also the fact that these technologies can be more effectively protected through patents compared to the other technologies \citep{cohen2000protecting}. 
Moreover, as coastal adaptation has public goods characteristics which may explain that private incentives for innovation ---and patenting--- can be relatively dampened. 
Patenting in adaptation related to water and infrastructure peaked in the 1980s but subsequently tapered off. 

\subsubsection{{Drivers of innovation in adaptation and mitigation}}
\black{Previous research has shown that innovation and patenting in green technologies respond to price signals and the size of the market for the technology \citep{popp2002induced, acemoglu2012DTC}. The market size of a technology can grow through an increased demand, for example induced by environmental disasters \citep{miao2014necessity} or regulatory pressure \citep{andreen2003water, kemp2000should}. These drivers may explain patterns in the time series data. }

Health- and water-related adaptation rose in the aftermath of the first regulatory initiatives by the US government to reduce the pollution of the air and water resources by industrial processes (e.g. the Clean Air Act in 1963 and Clean Water Act in 1972). \black{The majority of health-related CCATs during the 1960-1980s are air-pollution control technologies, and water-related adaptation technologies comprise many inventions for water treatment and pollution control (see \ref{SI:y02_definitions}).  
Innovation in pollution-control technologies is one response to regulatory pressure \citep{andreen2003water, kemp2000should} with an enhanced technological capacity for health- and water-related adaptation as a byproduct.}

\black{The rise in infrastructure adaptation can be associated with the Oil Crisis. 
Increased energy costs in response to that crisis were an important driver of energy-saving innovations \citep{popp2002induced, hassler2021directed}. 
Patents for infrastructure adaptation comprise many energy-saving insulation technologies, for example preserving thermal comfort in buildings or making power lines for energy transmission more robust (see \ref{SI:y02_definitions}).} 

\black{Innovation can be also triggered by an \emph{entrepreneurial state} \citep{mazzucato2011entrepreneurial} that actively engages in basic and applied research and creates markets for novel technological solutions. Many of the early inventions in low-carbon energy technologies during the 1950s and 1960s can be attributed to upcoming government-led initiatives in nuclear energy \citep{cowan1990nuclear}, but also in renewable energy technologies emerging from early US government initiatives from the Department of Energy and the US space program \citep{mazzucato2011entrepreneurial}. }

\subsubsection{Reliance on public support} 
Patenting in adaptation strongly relies on governmental support with over 40\% of patents since mid-2000s being linked directly or indirectly to government support (Fig. \ref{subfig:ts_share_green_gov_supp}). 
This is about 10\% higher than the value for average patents in the US \citep[cf.][]{fleming2019government}. Indirect adaptation and health-related adaptation show the highest levels of reliance on public support (Fig. \ref{subfig:ts_share_adap_gov_supp}). 
For many climate change mitigation technologies, by contrast, we observe a lower reliance on public support today compared to earlier periods. Especially clean energy and green ICT were heavily supported in the past, but have seen a significant private-sector take off. Mitigation technologies with insufficient market demand (e.g. CCS) show comparably high levels of public support as adaptation technologies. 

\black{The reliance on public support serves as an indicator of the stage of market development: if sufficient market demand for a technology exists, innovators have a commercial interest to develop these technologies and the reliance on public support is low. In contrast, if markets are underdeveloped (as in the case for adaptation and CCS), the public sector can play a critical role to stimulate innovation \citep[cf.][]{mazzucato2011entrepreneurial}. }

\black{Reliance on public support also includes the reliance on research that is funded by the government. As we shall see below in \ref{sec:science_base} and \ref{SI:scientificness}, adaptation technologies are more science-reliant than many other technologies. This contributes to the relatively higher reliance on public support of adaptation technologies (but also of science-reliant CCS), but the time trends suggest that this is not sufficient to explain this pattern. For example, for both clean energy and green ICT, the science reliance increased over time, but we observe a decreasing reliance on public support. Moreover, among the different categories of adaptation technologies, we also find that technologies like coastal, water-related, and infrastructure adaptation exhibit relatively higher shares of reliance on public support despite low levels of scientificness.}

\FloatBarrier
\subsection{The knowledge base of adaptation}
\label{sec:science_base}

To study the knowledge base of adaptation technologies, we combine data on patent citations, co-classifications \citep{hotte2021data}, and science citations \citep{marx2019reliance}. 
Citations from patents to science indicate the scientific origins of patented inventions \citep{meyer2000does, ahmadpoor2017dual}. Similarly, citations from patents to other patents describe technological base of patented inventions \citep{jaffe2019patent, verhoeven2016measuring}. 

\subsubsection{Reliance on science: Two clusters}

We find that adaptation technologies, as reflected by patents, can be grouped into two clusters: (i) science-intensive technologies (agriculture, health, and indirect adaptation); and (ii) engineering-based technologies (coastal, water, and infrastructure).

\begin{figure}[h]
    {\centering
        \caption{Science intensity of adaptation technologies}
    \label{fig:ts_scientificness_tech}

    \includegraphics[width=\textwidth]{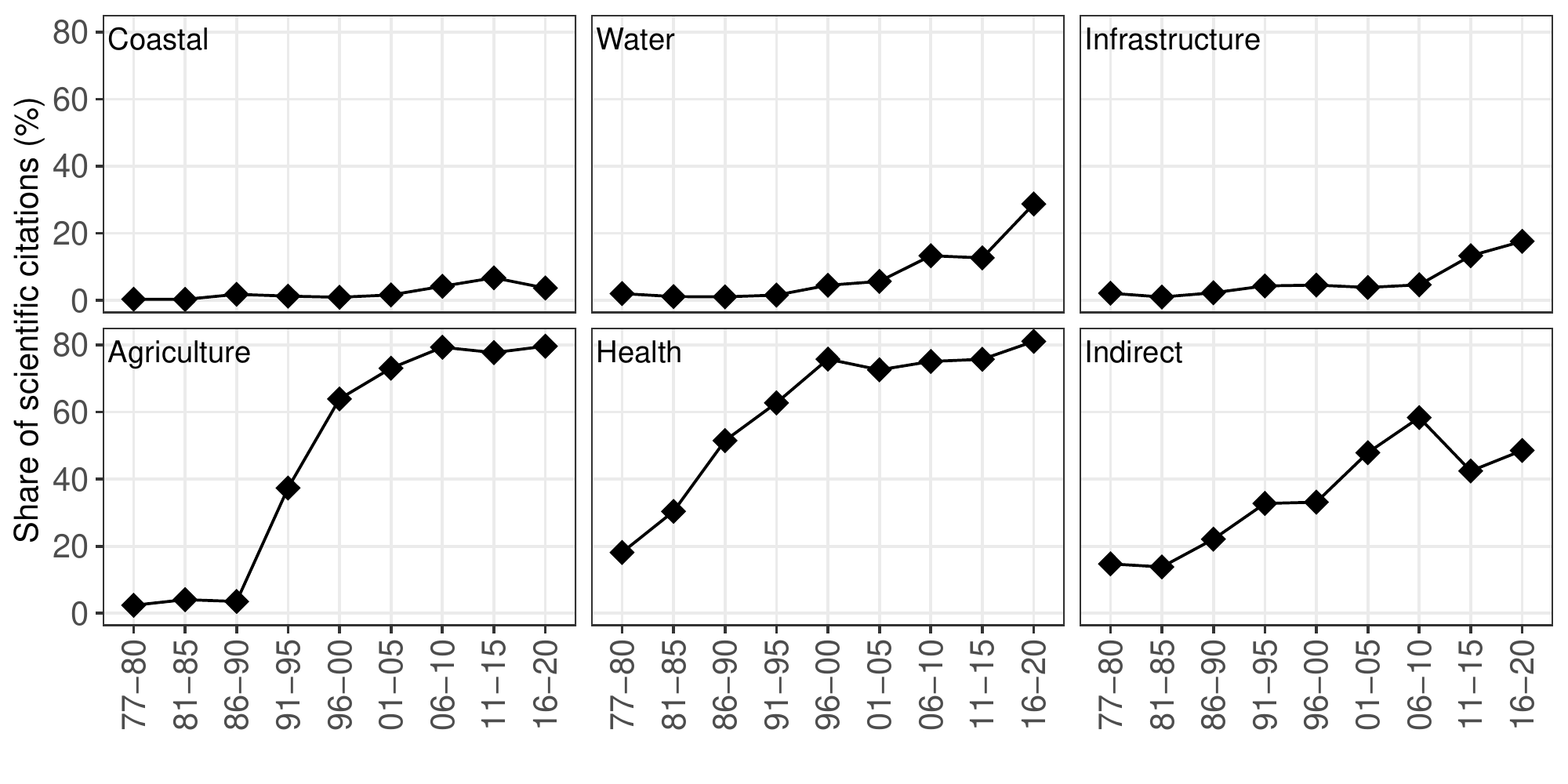}

    }
    \scriptsize
    Notes: Science-intensity of different adaptation patents measured by the share of citations to science in the number of total citations (sum of citations to other patents and scientific articles). 

\end{figure}
We measure the scientificness of adaptation technologies by the share of patent citations to science over the sum of citations to other patents plus citations to science. This ratio indicates to which extent a patent relies on science rather than applied technological development as encoded in patent citations \citep[see][]{hotte2021rise}.
The evolution of the CCATs' scientificness over time since 1976 is shown in Fig. \ref{fig:ts_scientificness_tech}. 
Coastal, water, and infrastructure adaptation technologies exhibit low shares of citations to science (0-5\%) while health, agriculture, and indirect adaptation are highly science-intensive (50-80\%). This reflects the idiosyncratic nature of different technologies. 
To be specific, science-intensive adaptation technologies include, for example, crops that are climate resilient, treatments for diseases that will become more prevalent in hotter temperatures, and complex early warning and monitoring systems. By contrast, engineering-based adaptation, which relies significantly less on science, includes technologies such as fixed construction to provide flood defense, cliff stabilization, water purification, and methods to strengthen the resilience of infrastructure.

The rise in share of citations to science for agriculture and health in the 1970s-1980s coincides with the rise of the US biotechnology. This period was characterized by many spin-offs from universities and public research laboratories that undertook innovation in basic necessities \citep{powell1996interorganizational, powell2005network}. 
Even within the engineering-intensive adaptation technologies such as water and infrastructure adaptation, we observe that these technologies became more science-intensive. We find that this phenomenon is related to the increased scientificness of chemistry-reliant water-conservation technologies (such as desalination, reverse osmosis), advances in material sciences for infrastructure adaptation, and increased interactions between developments in science-reliant solar photovoltaics with water and infrastructure adaptation, for example to supply energy for water treatment or heating and cooling in buildings.
 
Coastal adaptation, the smallest category in our sample, did not show an increase in its reliance on science. This is exceptional as an increasing reliance on science is a general trend in innovation during the second half of the twentieth century \citep{hotte2021rise}. This indicates that other knowledge sources rather than science are important for patented technologies in coastal adaptation, although interpretations must be made with caution due to the low number of patents. 



\subsubsection{Composition of the scientific base}

\begin{figure}[h]
    {\centering
        \caption{Composition of scientific knowledge base by scientific fields}
    \label{fig:reliance_on_wos6}
    \includegraphics[width=\textwidth]{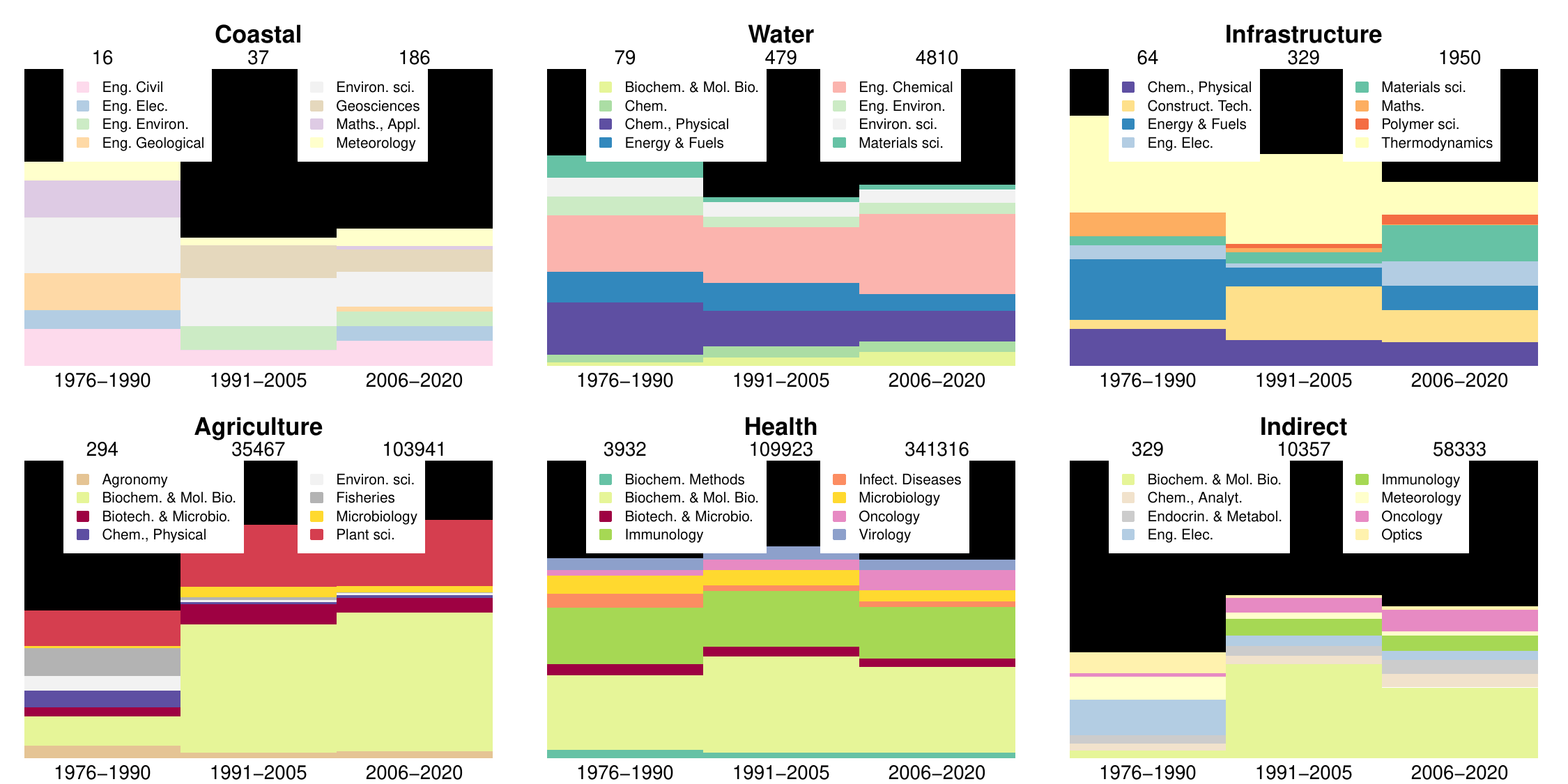}
}

\scriptsize
Notes: These figures show the 8 most often cited scientific fields (by Web-of-Science categories). The numbers on top of each bar indicate the number of papers cited by patents granted in the respective time period. The size of the colored fields in each bar plot indicates the share of citations that goes to the respective WoS field. Black color is used for the residuum of fields that are cited less often than the 8th most often cited field. 
\end{figure}

We studied the knowledge base of adaptation showing which fields of science are cited by adaptation technologies over time (Fig. \ref{fig:reliance_on_wos6}). 
This gives an idea of the scientific disciplines policymakers can support to strengthen innovation in adaptation technologies. \black{Complex technologies require a so-called \emph{absorptive capacity} to be effectively used and further developed \citep{cohen1990absorptive, caragliu2012impact, criscuolo2008novel}. In many environmental technologies, off-the-shelf solutions available on global markets require adaptive innovation to become useful under locally specific conditions \citep{popp2020international}. 
Hence, having expertise in scientific fields that are relevant for adaptation can spur the adoption, adaptation, and indigenous development of CCATs and ensure their maintenance. 
This can facilitate the efficient transfer of CCATs to regions where being exposed to climate risk, which is particularly urgent in many developing countries \citep{huenteler2016effect, adenle2015global, lema2016low}. These regions can stimulate adoption of adaptation technologies by investing in laboratories of local universities or public research institutions having relevant scientific understanding and thereby stimulate the transfer of adaptation skills to the local community.} 

Distinguishing between applied and basic research following \citet{persoon2020science}, we find that science-intensive CCATs (agriculture, health, and indirect adaptation) rely mostly on basic research, while adaptation technologies with a low science-intensity (coastal, water, and infrastructure) build to a higher extent on applied research.

Among the science-intensive CCATs, both health and agriculture largely build on biochemistry and molecular biology. Health adaptation further relies on immunology, oncology, and virology, while agricultural adaptation further relies on plant sciences. Indirect adaptation technologies which cover monitoring, assessment, and forecasting technologies rely on physics-related areas such as electrical engineering and optics, which form foundations of sensor and measurement technologies. Further, they build on biology-related areas such as biochemistry and immunology. Manual inspections of patents reveal that indirect adaptation technologies cover not only weather forecasting and monitoring technologies but also bioinformatics technologies for medicine and chemical assessment. \black{Therefore, university or public research laboratories in the field of biochemistry or molecular biology would be a good starting point for transferring many of the science-intensive adaptation technologies to the regions in need of such skills and knowledge \citep{adenle2015global}.} 

In engineering-based CCATs, applied sciences dominate. Coastal adaptation relies on several different types of engineering (civil, electric, environmental, and geological), but it also has weak linkages to some basic research of meteorology, maths, geosciences and environmental science. Water-related adaptation also relies on engineering but also basic research in chemistry, which is relevant for water conservation, filtration, recovery, and desalination that make use of chemical processes. The scientific knowledge base of infrastructure adaptation consists of material science, thermodynamics, construction, and electrical engineering, among other fields. \black{To sum up, to transfer the engineering-based CCATs to the regions in need, the role of laboratories in the engineering department of local universities will be particularly important, though basic science is also necessary in some fields. For example, in regions at high risk of sea level rise, civil engineers, and geologists in local universities may work together to efficiently adopt and advance technologies for coastal adaptation, and to adapt them to locally-specific conditions. Similarly, in regions where water adaptation is urgent, chemical engineers in the local universities may play a pivotal role in facilitating the adoption and further development of water adaptation technologies.}

\subsubsection{Composition of the technological base}

\black{A single patent can belong to multiple technology classes, reflecting a combinatory nature of knowledge creation \citep{ nelsonevolutionary}. Investigating co-classification patterns of adaptation patents can reveal technological capabilities other than Y02A that are needed to develop each type of CCAT. In addition, the co-classification patterns can be also interpreted as reflecting the promising fields of technological convergence with adaptation technologies \citep[e.g.][]{jee2019exploring}.}

\black{Therefore, organizations equipped with capabilities in fields frequently co-classified with Y02A can be understood as being in a competitive position in developing and exploiting adaptation technologies. Motivating these organizations, particularly in the private sector, to engage in the development 
of adaptation technologies can be a reasonable direction to spur innovation in climate change adaptation. In addition to encouraging the supply side, governments can also stimulate targeted foreign direct investments (FDI) or foreign licensing and connect these organizations with potential regions where demand exists, the regions being exposed to a higher risk of a certain type of climate change \citep{ferreira2020technology, popp2020international, saggi2002trade}. Targeted technology-transfer policy may not only stimulate the diffusion of environmentally related technologies, but also spur technological learning and indigenous innovation by local firms.}

\begin{figure}
    {\centering
        \caption{Co-classification of adaptation technologies}
    \label{fig:coclasses_1digit}
    \includegraphics[width=\textwidth]{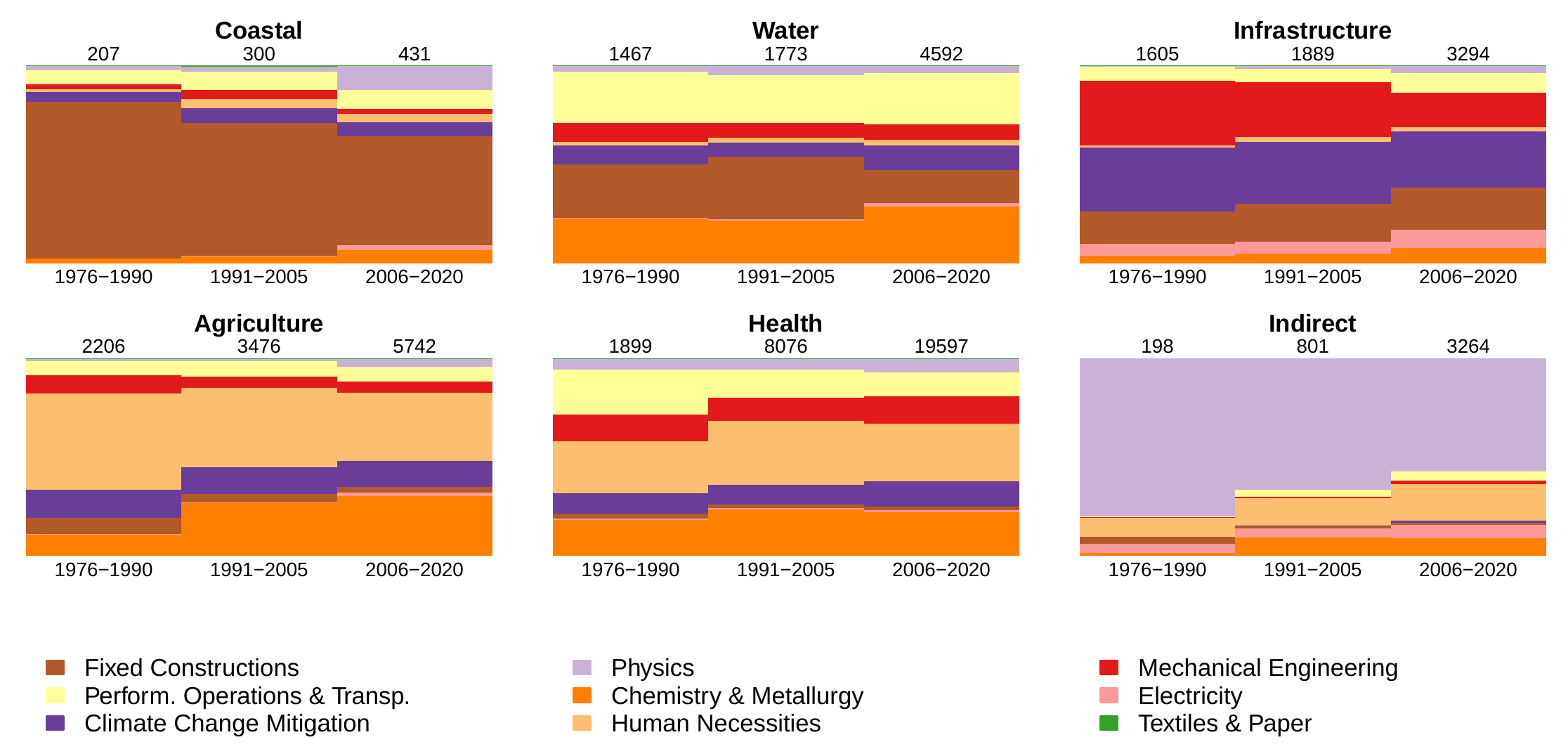}
}

\scriptsize
Notes: These figures show the co-classification of adaptation technologies at the CPC section level (1-digit). 
The numbers on top of each bar indicate the number of patents granted in each sub-period. Note that the bar plots rely on the number of co-classifications. Patents that serve multiple adaptation purposes, i.e. are classified into multiple adaptation technology types, are double-counted. The size of the colored fields in each bar plot indicates the share of co-classifications for different subgroups by adaptation technology type. 
\end{figure}

\black{Fig. \ref{fig:coclasses_1digit} shows the overall patterns of co-classification for each type of CCAT. For example, we can see that the vast majority of coastal adaptation patents are co-classified as fixed constructions technology.\footnote{\black{Coastal adaptation significantly relies on solutions that are difficult to patent as well, such as mangrove reforestation and nature-based solutions. We should note that Fig. \ref{fig:coclasses_1digit} includes a bias towards coastal adaptation solutions that are patentable, rather than the hard to be patented solutions.}} The results imply that firms with fixed construction engineering skills are in a good position to develop and utilize coastal adaptation technologies. Targeted government support on these firms to motivate their investment in coastal adaptation technologies and to match them with regions with high risk of sea level rise would play an important role in stimulating innovation in coastal adaptation.}

\black{Many indirect adaptation patents are co-classified as physics (see Fig. \ref{fig:coclasses_1digit}). In-depth analysis with further technological details (see \ref{fig:coclasses_4digit_8}) shows that this is due to technological interdependencies between indirect adaptation and applied physics including measurement, detection, and prediction technologies. Therefore, to stimulate innovation in indirect adaptation, governments can incentivize firms with advanced skills in measurement, detection, and prediction to invest in indirect adaptation technologies, as well as connect these firms to regions where precise, timely sensoring and forecasting of climate disaster are critical.} 

Fig. \ref{fig:coclasses_1digit} also shows the extent to which different categories of adaptation patents are labeled as mitigation patents, indicated by purple color. 
The next section explores this duality in more detail.

\FloatBarrier
\subsection{Complementarities with mitigation}
\label{sec:mitigation_adaptation}
We next focus on complementarities between adaptation and mitigation technologies to inform technology-choices that help achieve climate change mitigation and adaptation at the same time. 
We use two different approaches: (1) analyzing patents that are co-classified as adaptation and mitigation technologies to identify `dual purpose' technologies, and (2) examining the extent to which adaptation and mitigation technologies rely on similar technological and scientific knowledge (i.e., cite the same patents and papers). \black{The knowledge base similarity of adaptation and mitigation technologies helps understand how mutual knowledge spillovers between adaptation and mitigation can be stimulated. For example, public support may be directed towards the fields in which both adaptation and mitigation rely on.}

\subsubsection{Adaptation technologies with mitigation co-benefit}
\begin{table}[!h]
\centering
{

\begingroup\scriptsize
\caption{Overview statistics of dual purpose patents}
\label{tab:overview_MA_patents}
\small
\begin{tabular}{p{4cm}p{1.8cm}p{1.8cm}p{1.8cm}p{1.8cm}p{1.8cm}}
  \hline
\hline
 & Total patents & Share dual purpose & Citing/total patents & Share gov. supported\\ 
  \hline
      \multicolumn{5}{l}{\textbf{\emph{Dual purpose adaptation technologies}}}\\
        \hline
        Coastal &  71 & 0.08 & 0.11 & 0.30 \\ 
  Water & 926 & 0.20 & 0.25 & 0.32 \\ 
  Infrastructure & 3276 & 0.70 & 0.10 & 0.17 \\ 
  Agriculture & 1768 & 0.22 & 0.21 & 0.23 \\ 
  Health & 3668 & 0.22 & 0.19 & 0.20 \\ 
  Indirect &  39 & 0.01 & 0.54 & 0.69 \\ 
\hline
Gov. support (No) & 6440 & 0.27 & 0.10 &  \\ 
  Gov. support (Yes) & 1657 & 0.17 & 0.41 &  \\
\hline
All & 9645 & 0.26 & 0.23 & 0.20 \\ 
      \hline
  \hline
  \multicolumn{5}{l}{\textbf{\emph{Dual purpose mitigation technologies}}}\\
  \hline
Buildings & 3033 & 0.07 & 0.09 & 0.16 \\ 
  CCS & 429 & 0.08 & 0.44 & 0.46 \\ 
  Green ICT &  15 & 0.00 & 0.67 & 0.36 \\ 
  Energy & 920 & 0.01 & 0.24 & 0.31 \\ 
  Production & 1356 & 0.02 & 0.25 & 0.23 \\ 
  Transport & 2821 & 0.03 & 0.12 & 0.14 \\ 
  Waste & 1071 & 0.05 & 0.24 & 0.27 \\ 
\hline
Gov. support (No) & 6440 & 0.02 & 0.10 &  \\ 
  Gov. support (Yes) & 1657 & 0.02 & 0.41 &  \\
\hline
All & 9645 & 0.02 & 0.29 & 0.20 \\ 
\hline
\end{tabular}
\endgroup
}

\footnotesize \justify
Notes: This table summarizes the subsets of dual purpose patents, i.e. patents that are simultaneously classified as adaptation and mitigation technology. The upper part of the table shows these patents from the angle of adaptation, the lower part from the angle of mitigation technologies. 
Column \textit{Share dual purpose} shows the share of dual purpose patents in all patents. Column \textit{Citing/total patents} shows the ratio of patents that cite at least one scientific paper over the number of all patents. Column \textit{Share gov. supported} shows the share of patents that benefited from governmental support. 
The rows \textit{Gov. support: Yes} and \textit{No} show statistics for the subset of patents that do and do not rely on public support, respectively. Sum of the number patents in each sector is not exactly same with the number of all patents because a patent can be classified into multiple sectors at the same time. 
\end{table} 

\black{Starting off with co-classifications, we find that many adaptation patents except for those in indirect adaptation include a significant proportion of dual purpose patents helping in not only adaptation but also mitigation (purple bars in Fig. \ref{fig:coclasses_1digit}). 
In total, 26\% of adaptation patents are co-classified as mitigation patents, showing that more than a quarter of adaptation technologies have the potential to be used in both adaptation and mitigation areas} (Table \ref{tab:overview_MA_patents}). The highest overlap is in infrastructure adaptation where 70\% of the patents are co-classified as mitigation technologies. \black{For example, thermal insulation in buildings achieves both adaptation and mitigation purposes: it preserves thermal comfort during extreme temperature events, but it may also help reduce energy consumption and associated emissions. This is an example of how maladaptation relying on the intensified use of air-conditioning to cope with heatwaves can be avoided \citep{barnett2010maladaptation}. Other illustrative examples are extreme weather resistant electricity grids that rely on insulation technologies that help reduce energy losses during the transmission through the grid, or integration of production and use of renewable energy into buildings for heating and cooling purposes.} 

For health-, agriculture-, and water-related adaptation, roughly 20\% of patents simultaneously serve mitigation purposes (Table \ref{tab:overview_MA_patents}). \black{Co-benefits in health adaptation arise for example from clean transportation that reduce emissions. This represents a preventive intervention improving public health as air-pollution control helps prevent respiratory and cardivascular diseases. Research has shown that these diseases increase the vulnerability to heatwaves and some infectious diseases \citep{harlan2011climate}, including Covid-19 \citep{domingo2020effects}}. 

In agriculture, we find technologies that improve the climate resilience of plants can simultaneously sequester carbon. Some technologies that contribute to an improved handling of bio-related waste or energy efficiency of greenhouses can simultaneously be used in cooling systems for food storage. 

In addition, some adaptation technologies used for water treatment, purification, and desalination also help reduce emissions in wastewater and solid waste treatments. 
\black{\citet{barnett2010maladaptation} mentioned energy-intensive desalination as an example of emission-increasing maladaption. Our analysis shows that mitigation-friendly alternatives exist, combining renewable energy with desalination.}

By contrast, the occurrence of dual purpose technologies is relatively weak in coastal (8\%) and indirect adaptation technologies (1\%). 

When examining the degree to which mitigation patents can be co-classified as adaptation patents, we find that CCS, clean buildings, and waste management related mitigation technologies include 8
By the number of patents, clean transportation, efficient production, and low-carbon energy patents have significant co-classification with adaptation. However, due to the large number of patents in these categories the share of co-classification is low, ranging between 1-3\%. 

\black{The fact that some adaptation technologies bear mitigation co-benefits does not tell us much about the climate impact of the remainders beyond examples mentioned in the literature on maladaptation. 
We cannot say ---based on our analysis--- whether mitigation technologies that are not co-classified as adaptation have a negative or positive impact on the economy's climate resilience. While not judging whether adaptation and mitigation are complements in general, we show that some adaptation-mitigation options are complementary. Complementarity may be a matter of technology choice and our analysis identifies areas that are promising to achieve adaptation-mitigation co-benefits.}

\black{Although our analysis shows the technological potential to achieve both adaptation and mitigation goals, the absence of co-benefits does not necessarily imply an inferior technology choice. Other factors such as competing policy objectives, economic constraints, different time horizons, and locality of events may constrain the set of available technology options. For example, health-related adaptation technologies to cope with risks from vector borne diseases are urgent in some developing countries although disconnected from any mitigation technology.
For nuclear energy, in some countries, political objectives to achieve short-term mitigation may weigh higher than the long-term resistance to climate change. At least in the short term, adaptation co-benefits of nuclear energy are absent and there is good reason to believe that these technologies rather undermine than strengthen the vulnerability against extreme climate shocks \citep{hanski2018assessing, jordaan2018resilience}. Nevertheless, short term mitigation benefits of this technology are strong and ---assuming a positive mitigation impact--- it also contributes to adaptation in the long run if it helps reduce the impact of climate change.}



\subsubsection{Potential knowledge spillovers between mitigation and adaptation}

\begin{figure}
{    \centering
\caption{Technological and scientific similarity among CCATs and CCMTs}
\label{fig:similarities}

    \begin{subfigure}{.45\textwidth}
    \centering
    \includegraphics[width=1\textwidth]{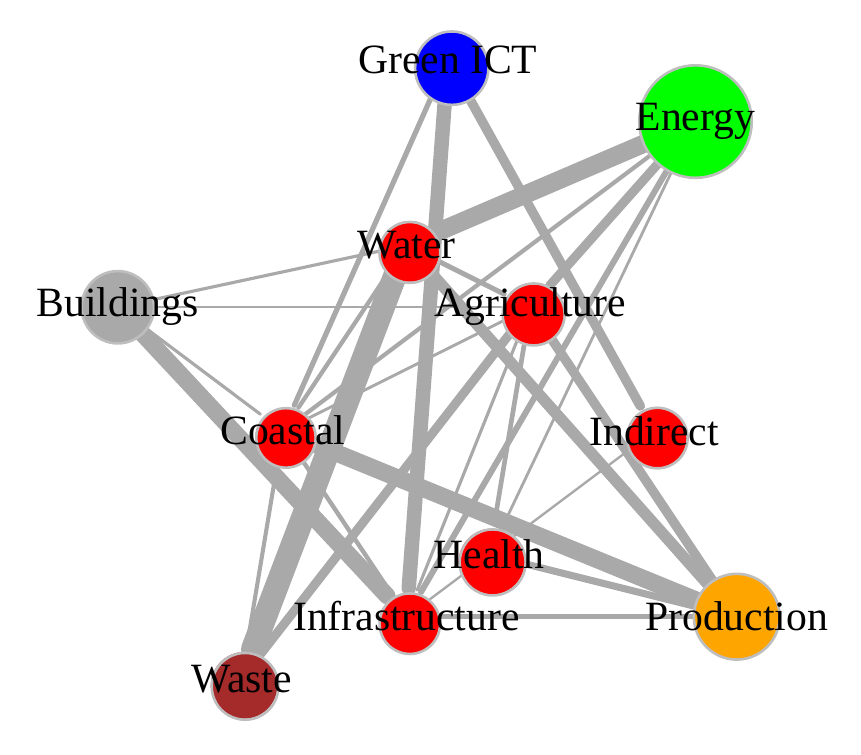}
    \includegraphics[width=0.9\textwidth]{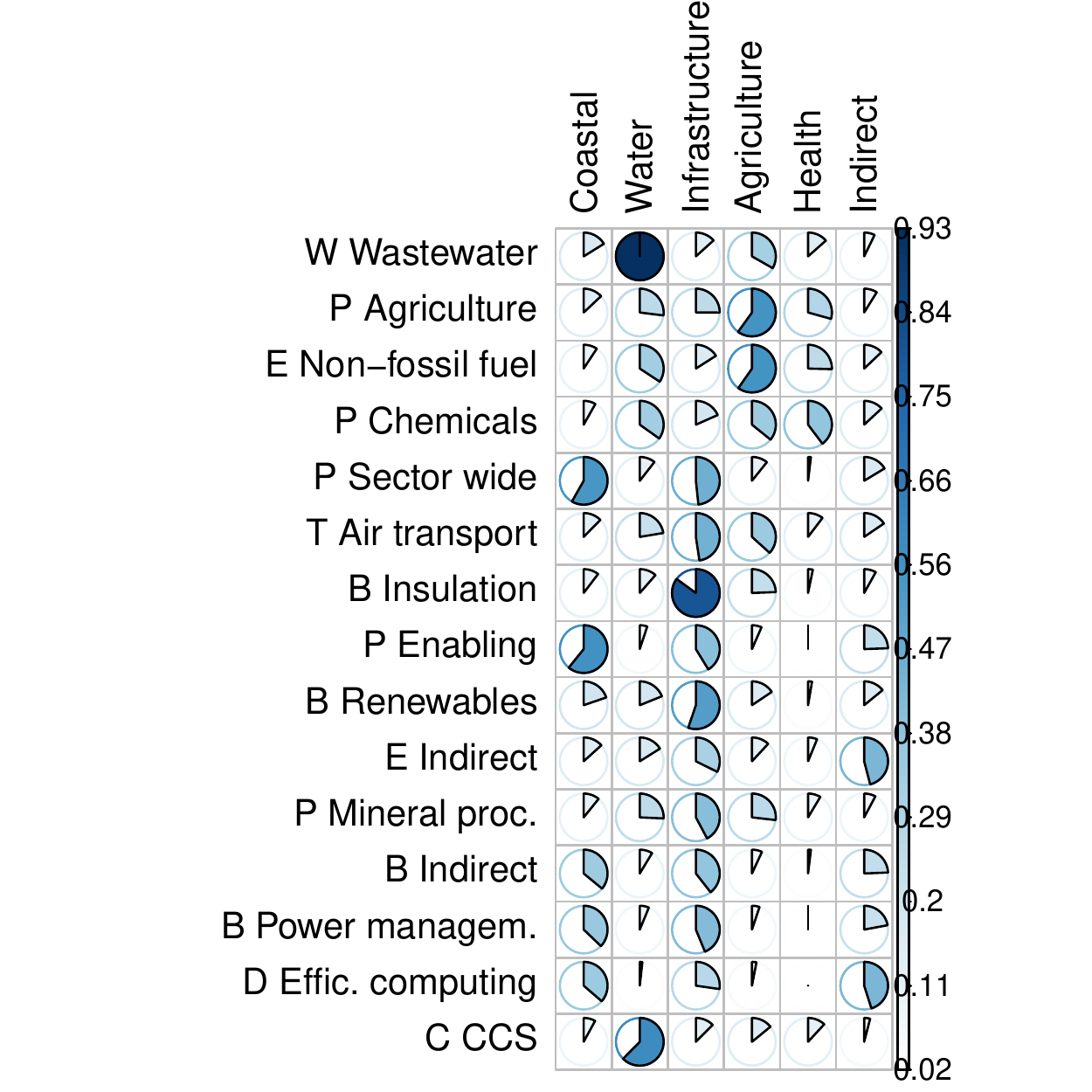}
    \caption{Technological similarity}
    \label{subfig:similarity_cpc_MA_complements}
    \end{subfigure}
            \begin{subfigure}{.45\textwidth}
    \centering
    \includegraphics[width=1\textwidth]{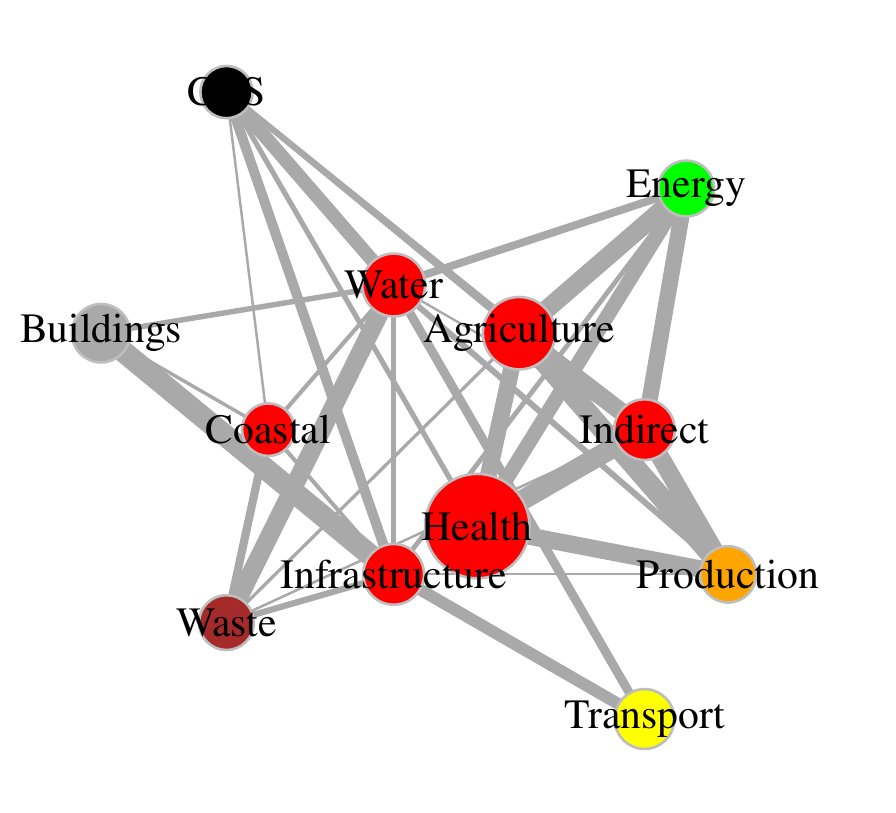}
    \includegraphics[width=0.9\textwidth]{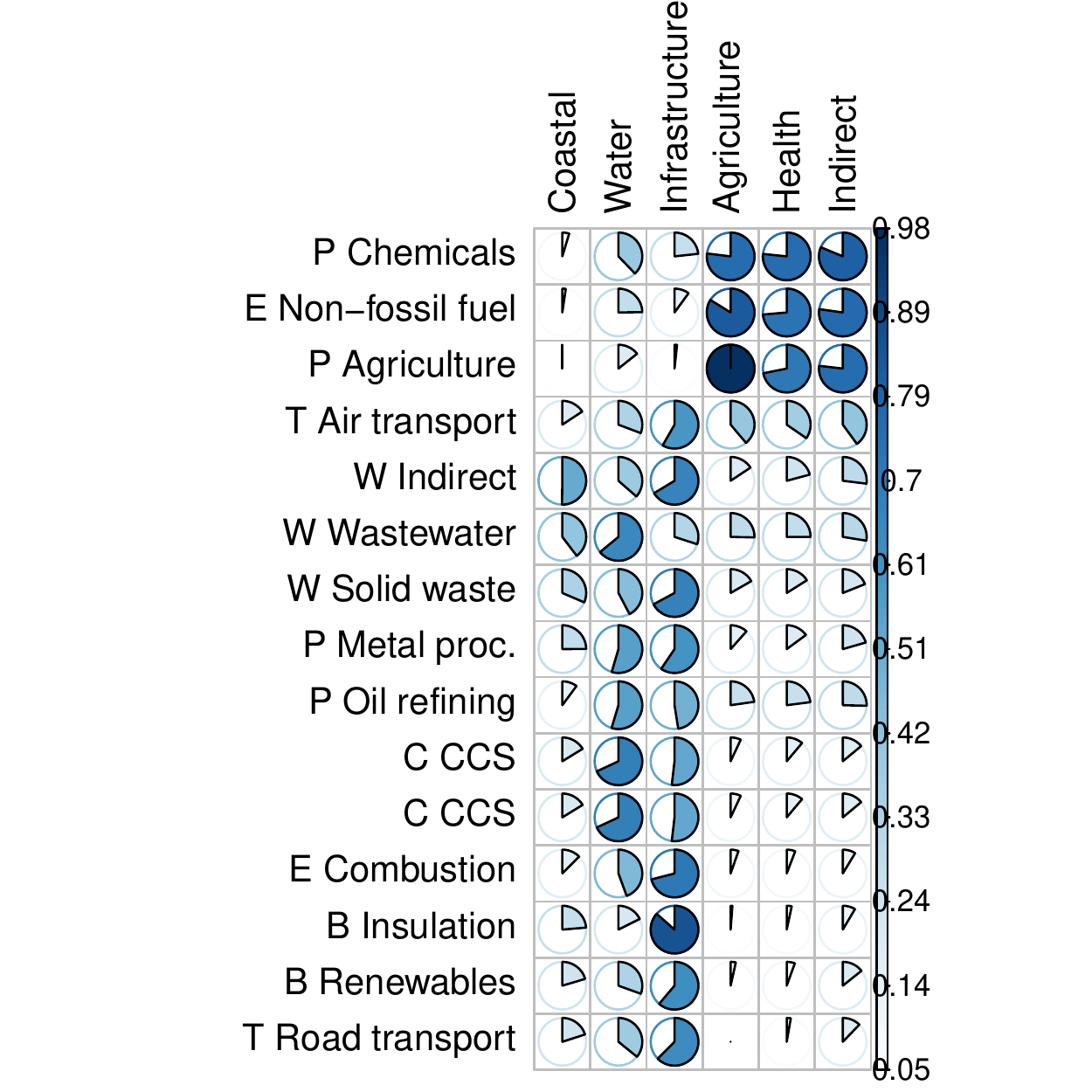}
    \caption{Scientific similarity}
    \label{subfig:similarity_wos_MA_complements}
    \end{subfigure}

}

    \footnotesize 
   Notes: These figures illustrate technological similarities of different types of adaptation and mitigation technologies that are complements, i.e. simultaneously classified as adaptation and mitigation technologies. 
   The figures are based on shares of (a) citations to CPC 4-digit technology classes and (b) citations to scientific fields (WoS). 
   The upper two figures show similarity networks. A link between a pair of technologies indicates the cosine similarity of their references to scientific fields and technology classes, respectively. For clarity only the most significant links are shown. 
   The widths of connecting edges are proportional to the degree of similarity and the node sizes are proportional to the number of patents. 
   The node colors indicate the 4-digit technology class (i.e., red for adaptation, gray for buildings, black for CCS, blue for green ICT, green for energy, orange for production, yellow for transport, and brown for waste). The lower two figures illustrate the numerical values of the cosine similarity of adaptation (columns) and mitigation (rows) technologies at the 6-digit level. The letters in the beginning indicate the type of mitigation technology (i.e., B for buildings, C for GHG disposal, D for green ICT, E for energy, P for production, T for transport, and W for waste).  
\end{figure}


\black{We investigate the extent to which different mitigation and adaptation technologies build on a common knowledge base. To identify domains of shared knowledge, we analyze similarities of the technological and scientific knowledge base for pairs of different adaptation and mitigation technologies. 
Previous research has shown that similarities enable knowledge spillovers across technologies at the organizational, regional, and national level, and they are an indicator of absorptive capacity as it is easier for firms, industries, and countries to adopt a new technology if the adopter has pre-existing relevant knowledge \citep{cohen1990absorptive, caragliu2012impact, criscuolo2008novel}. 
This also matters for policy: if two technologies build on the same knowledge sources, R\&D policy may focus on these areas to support the development of both technologies at the same time. }

In Fig. \ref{fig:similarities}, we illustrate knowledge similarities through network plots and correlation charts. Similarities are measured via backward citation patterns: two technologies are more similar if they rely more on common sources of knowledge. This is measured by the cosine similarity based on shares of citations to CPC 4-digit technology classes (Fig. \ref{fig:similarities}(a)) and citations to scientific Web of Science (WoS) fields (Fig. \ref{fig:similarities}(b)). 

\black{The upper two figures show similarity networks. A link between a pair of technologies indicates the cosine similarity of their references to scientific fields and technology classes, respectively. For clarity, only the most significant links are shown.\footnote{We use the median weight of connecting links as significance threshold and show only those links whose weight is larger than that.} The widths of connecting edges are proportional to the degree of similarity and the node sizes are proportional to the number of patents. 
The node colors indicate the 4-digit technology class (i.e., red for adaptation, gray for buildings, black for CCS, blue for green ICT, green for energy, orange for production, yellow for transport, and brown for waste).}
  
\black{The lower two figures illustrate the numerical values of the cosine similarity of adaptation (columns) and mitigation (rows) technologies at the 6-digit level. The letters in the beginning indicate the type of mitigation technology (B for buildings, C for GHG disposal, D for green ICT, E for energy, P for production, T for transport, and W for waste).} 
\black{Our similarity analysis shows:} 

(1) Citing similar patents, mitigation technologies for energy efficiency in buildings and green ICT have a similar technological knowledge base to infrastructure-related adaptation technologies. 
Technologies that reduce transmission losses and improve the energy efficiency of ICTs rely on similar technological knowledge as technologies that strengthen the resilience of physical infrastructure to extreme weather events. The same holds for insulation, efficient heating, and renewable energy in buildings.  

(2) Clean energy, especially clean combustion and bio-fuels, exhibits strong scientific similarities with science-intensive adaptation technologies such as agriculture, health, and indirect adaptation. This is particularly due to their common reliance on chemistry (see Section \ref{sec:science_base} and for more detail Fig.  \ref{fig:MA_cosine_similarties_plot_76_06}-\ref{fig:family_based_similarity_networks_complements}, \ref{fig:SI_science_base_MA_complements_mitigation_6} in the Supplementary Material). 

(3) Water-related adaptation technologies exhibit a high degree of scientific similarity with clean industrial processing technologies for metal and oil, with waste treatment, and CCS. This can be explained by their joint reliance on chemistry. 
We also observe a high potential for scientific and technological knowledge spillovers between water adaptation and clean energy that mostly arise from interactions with non-fossil fuels and renewables. Our data shows that examples of energy intensive water treatment technologies like desalination explicitly make use of photovoltaics, which explains their reliance on the same science (see \ref{SI:y02_definitions} and for more detail Fig. \ref{fig:MA_cosine_similarties_plot_06_20}-\ref{fig:family_based_similarity_networks_complements} in the Supplementary Material).


(4) We observed the rise in mitigation technologies for clean production that have adaptation co-benefits (see orange node in Fig. \ref{fig:similarities} and \ref{fig:MA_cosine_similarties_plot_06_20}-\ref{fig:SI_science_base_MA_complements_mitigation_6}). 
Not surprisingly, the spillover potential is highest in between adaptation and mitigation in agricultural production. In addition, we observe a large potential for technological knowledge spillovers between enabling technologies in production and various fields of adaptation. This suggests that there is a high potential to harness knowledge spillovers and to realign efforts to mitigate emissions in production processes with adaptation goals. 

\black{This analysis suggests that many adaptation and mitigation technologies in various domains share a common knowledge base. 
The reliance on similar technological and scientific knowledge suggests that R\&D investments in one area have positive side effects on another. Economically, the existence of positive knowledge spillovers is a justification for higher levels of public support, as the social returns of these investments exceed those from investments in technologies that show a lower spillover potential \citep{aldieri2019environmental}.} 

\FloatBarrier

\section{Discussion} 
\label{sec:discussion}
Despite the urgency of climate change and the substantial long-term economic benefits of adaptation \citep{tall2021enabling}, the study of innovation in adaptation has attracted relatively little scholarly attention \citep{popp2019environmental, dechezlepretre2020invention} and markets for adaptation technologies seem underdeveloped given their benefits \citep{dechezlepretre2020invention}.
However, this is likely to change: the requirement of countries to disclose their adaptation plans under the Paris Agreement \citep{lesnikowski2017does, berrang2019tracking}, increasing awareness of firms' climate risks and efforts by regulators to make risk disclosures mandatory will incentivize the public and private sector to take action towards adaptation \citep{goldstein2019private, smith2021climaterisk}. 
This study offers the first systematic analysis of adaptation technologies and their knowledge base addressing three questions. 

First, we asked: \emph{To what extent have these technologies been developed, and which were the drivers of innovation?} 
We find that patenting in most adaptation technologies did not increase substantially over the past decades, with the exception of health-related and indirect adaptation. 
\black{Historically, we observed several phases of increased inventive activity, especially since the late 1960s and during the Oil Crisis in the 1970s. The 1960s were the starting date of many environmental initiatives including regulatory measures such as the Clean Air Act from 1963 and Clean Water Act from 1972. As discussed above, many adaptation technologies, especially in water and health, interacting with pollution control bear a positive externality improving the environmental quality. Despite not providing causal evidence, 
the rise in certain water and health adaptation technologies we observe might be a byproduct of environmental regulatory policy.} 

\black{Similarly, our analysis revealed that many solutions for adaptation, especially in infrastructure and agriculture, have the potential to simultaneously improve energy efficiency. 
Many other technologies integrate off-grid renewable energy into their processes, for example for desalination, food processing and conservation, or cooling and heating in buildings. 
High prices for fossil fuel energy during the Oil Crisis stimulated investments in energy efficiency and the demand for alternative energy solutions, which offers one explanation for the rise adaptation, again as a byproduct of the seemingly unrelated energy price.}

Second, we addressed the question: \emph{How can governments support the development and adoption of these technologies?} 
\black{Our analysis has further revealed that public R\&D support may be supportive for early-stage technological development. This is particularly important for science-intensive technologies, as private sector incentives to engage in basic research with uncertain returns are limited.} Agriculture, health, and indirect adaptation technologies are highly science-intensive, while adaptation for coastal defense, infrastructure, and water is rather engineering-based. 

\black{Analyzing the scientific base of adaptation, we have further discussed that science-intensive CCATs rely more heavily on basic rather than applied sciences. 
This gives insights into policies for transferring adaptation technologies to regions in need. Local universities and public research institutions equipped with relevant scientific knowledge base (e.g., biochemistry and molecular biology for science-intensive CCATs) can be key actors in facilitating technology transfer, as they contribute to the regional absorptive capacity for science-intensive technologies. 
An analysis of co-classification patterns of adaptation patents helps identify organizations with complementary technological capabilities, which can be used to develop and exploit different adaptation technologies, for example firms with construction skills for coastal adaptation. Above, we discussed directions in which the government should provide targeted support for both supply and demand of adaptation solutions to stimulate innovation in climate change adaptation.}



Finally, we wanted to find out: \emph{How do technologies for adaptation interact with climate change mitigation?} 
From a technological perspective, climate change mitigation and adaptation are complements: on average, 26\% of adaptation technologies also help in mitigation. In some sub-fields such as infrastructure-adaptation, the complementarities are particularly large, with 70\% of adaptation patents simultaneously contributing to climate change mitigation. Well-designed policy may exploit and strengthen these complementarities to ensure that climate change technologies serve the twin goals of adaptation and mitigation. 

\black{Adaptation-mitigation co-benefits have been recognized in many adaptation case studies \citep{kabisch2017nature, sharifi2021co, berry2015cross}. Our analysis shows that this can be also seen systematically in aggregate data. This enables a systematic understanding of the drivers of innovation behind adaptation and shows many examples of how adaptation and mitigation efforts can be aligned. 
We have also seen that adaptation technology development often came as a byproduct of other economic trends. Identifying complementarity with other larger technological developments, for example in artificial intelligence and biotechnology, may help to make R\&D for adaptation more effective. 
Furthermore, systematic analyses of technological overlaps of adaptation with response strategies to major shocks such as Covid-19, the Ukraine war, or financial crises can also mobilize additional financial resources to create a resilient economy.}


\section{Conclusion}
\black{In this paper, we have taken stock of the current technological frontier of adaptation technologies. We have shown that ---compared to mitigation--- innovation in the field of adaptation has not yet taken off. In the analysis, we have identified and discussed major drivers of innovation in adaptation such as responses to regulation and shocks in the market, but we also highlighted a prominent role of the government stimulating the development of these technologies. }

\black{Our analysis has further shown how governments can effectively stimulate the development and adoption of technologies through targeted investments in scientific and technological capacities, and we discussed how this can help enable technology transfer to countries where adaptation needs are high.} 

\black{Finally, we addressed the nexus between climate change mitigation and adaptation and have shown that ---from a technological perspective--- adaptation and mitigation efforts may be complementary. However, this may be a matter of technology choice and our analysis may provide guidance on how these choices can be made to achieve mitigation and adaptation objectives at the same time. }

This study is limited to the technological frontier of adaptation as reflected in patent data. Although the granted patents capture inventions that have high (perceived) market value, patent data as a measure of innovation has well-documented limitations \citep{OECD2009} being biased towards the technological frontier solutions and being silent about other aspects such as nature-based or behavioral solutions. A promising avenue for future research is to develop measures for these other solutions of adaptation that can be systematically compared to the technologies analyzed in this paper. This would help understand the multiple trade-offs and synergies among different solutions for adaptation and their interaction with mitigation, which is highly relevant to address the climate challenge in an efficient way.


\section{Data availability}
The research data including R-scripts used for the data compilation and empirical analysis are publicly available under a CC-BY-4.0 license to ensure the full reproducibility of the results and re-use \citep{hotte2021data_adapt}. The data can be downloaded here: \url{https://doi.org/10.4119/unibi/2958327}. 

\section{Acknowledgments}
The authors want to thank Sugandha Srivastav who significantly contributed to an earlier version of this article, in both intellectual and practical ways. Further gratitude is owed to Anton Pichler and Fran\c{c}ois Lafond whose work on an earlier project contributed significantly to the methodological basis of this work. 
The authors also want to thank Matthias Endres, Peter Persoon, Vilhelm Verendel and their colleagues from the Institute for New Economic Thinking (INET), the Oxford Martin Pogramme on Technological and Economic Change (OMPTEC), and Future of Work for helpful feedback. Gratitude is owed to Elizabeth Champion for her proofreading assistance.  
K.H. acknowledges support from OMPTEC and Citi. S.J. acknowledges support from Basic Science Research Program through the National Research Foundation of Korea (NRF) funded by the Ministry of Education (2020R1A6A3A03037237). 
%
\section{Author contributions}
K.H. developed the research idea, study design, visualized the results, and wrote the initial draft. K.H. and S.J. compiled and analyzed the data. S.J. validated the results. All authors contextualized the results, reviewed, and edited the manuscript.

\section{Competing interests statement}
The authors do not have any competing interests to declare. 

\printbibliography

\newpage
\FloatBarrier

\appendix
\renewcommand{\appendixname}{Appendix}
\renewcommand{\thesection}{A.\arabic{section}} \setcounter{section}{0}
\renewcommand{\thefigure}{A.\arabic{figure}} \setcounter{figure}{0}
\renewcommand{\thetable}{A.\arabic{table}} \setcounter{table}{0}
\renewcommand{\theequation}{A.\arabic{table}} \setcounter{equation}{0}

\section{Y02A classes and definitions}
\label{SI:y02_definitions}
{This list is downloaded from:  \url{https://worldwide.espacenet.com/classification?locale=en_EP#!/CPC=Y02A} [April 2021].}

\begin{figure}[H]
    \centering
    \includepdf[pages={1},width=0.9\textwidth]{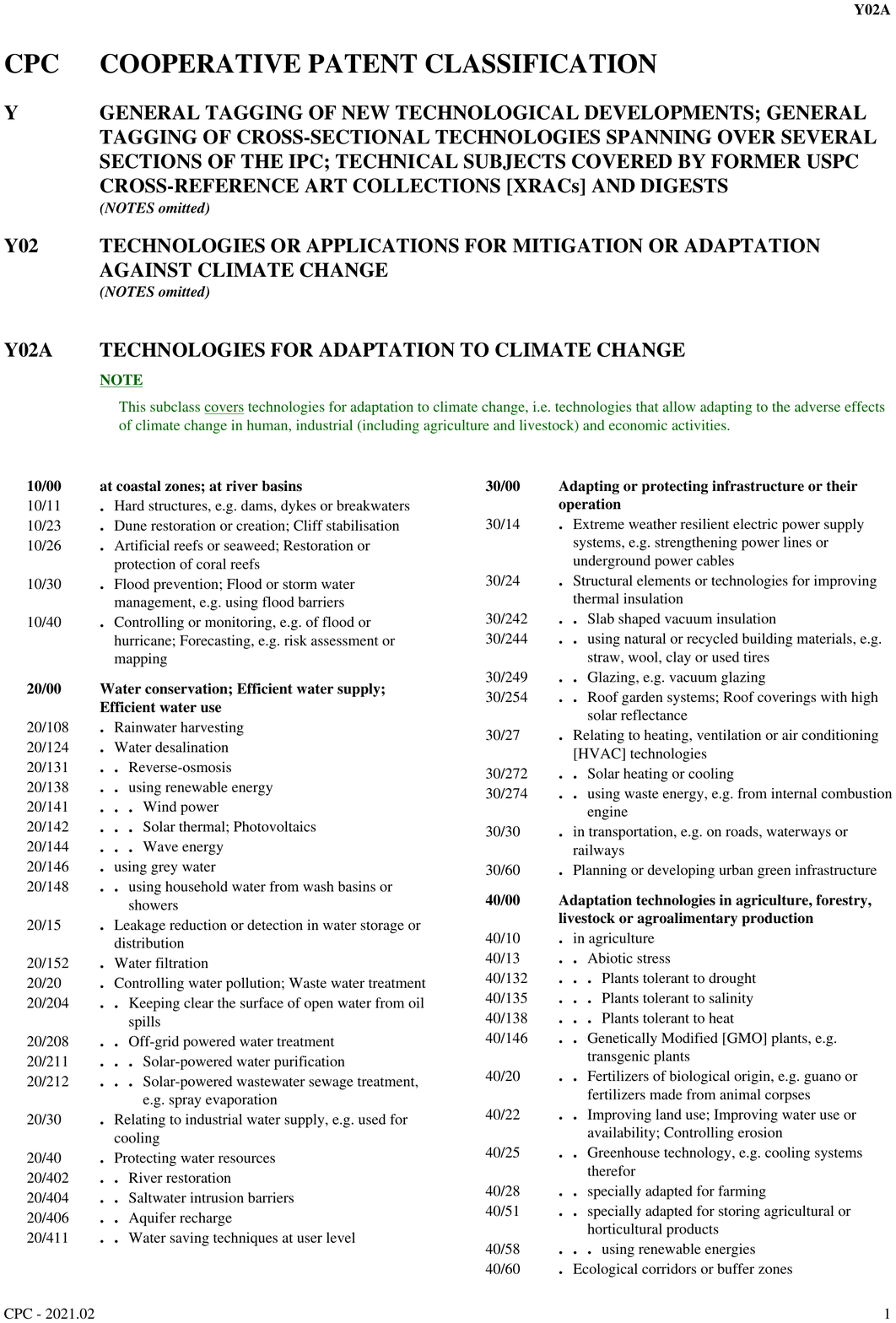}
\end{figure}
\newpage
\begin{figure}[H]
    {\centering
    \includepdf[pages={2},width=0.9\textwidth]{inputs_natCC/cpc-scheme-Y02A.pdf}
}

\end{figure}

\newpage
\FloatBarrier
\section{Additional results}
\subsection{Summary statistics of adaptation and mitigation patents}

In Table \ref{tab:overview_adaptation_patents}, we summarize the adaptation patents (at the DOCDB family level) granted over the full time horizon covered by our analysis. We have a relatively low number of patents in coastal adaptation (857) and the largest number of patents in health adaptation (16,363). We also report the number of patents that make at least one citation to science, the share of patents that cite to science, the number of scientific citations, average number of citations made by patents, average number of citations made by citing patent and the share of patents that are reliant on governmental support. 
Overall, we find that patents relying on public support tend to be more science intensive (i.e., exhibit a higher share of science reliant patents and make more citations to science than others). 
\begin{table}[ht]
{\centering
\begingroup\scriptsize
\caption{Overview of adaptation adaptation technologies}
\label{tab:overview_adaptation_patents}

\begin{tabular}{p{2.5cm}p{1.45cm}p{1.45cm}p{1.45cm}p{1.45cm}p{1.45cm}p{1.45cm}p{1.45cm}}
\hline
Technology & Total patents & Citing patents & Citing/total patents & Scientific citations & Citations per patents (all) & Citations per patents (citing) & Share Gov. support \\ 
  \hline
Coastal & 857 & 62 & 0.07 & 328 & 0.38 & 5.29 & 0.18 \\ 
  Water & 4678 & 720 & 0.15 & 14173 & 3.03 & 19.68 & 0.22 \\ 
  Infrastructure & 4671 & 421 & 0.09 & 3828 & 0.82 & 9.09 & 0.16 \\ 
  Agriculture & 8089 & 2482 & 0.31 & 149341 & 18.46 & 60.17 & 0.23 \\ 
  Health & 16363 & 9331 & 0.57 & 490310 & 29.96 & 52.55 & 0.39 \\ 
  Indirect & 2978 & 1749 & 0.59 & 72759 & 24.43 & 41.60 & 0.58 \\ 
   \hline
All & 37341 & 14681 & 0.30 & 730739 & 12.85 & 31.40 & 0.31 \\ 
\hline
Gov. support (No) & 21706 & 5717 & 0.26 & 158897 & 7.32 & 27.79 &  \\ 
  Gov. support (Yes) & 9853 & 6911 & 0.70 & 480611 & 48.78 & 69.54 &  \\ 
   \hline
\end{tabular}
\endgroup
}

\footnotesize
Notes: This table summarizes the characteristics of patents classified as climate change adaptation technologies. The categories are distinguished at the 6-digit CPC level. 
The column \textit{Citing patents} shows the number of patents that rely on science, i.e. make at least one citation to the scientific literature.  
The row entry \textit{All} corresponds to total numbers for columns \textit{total patents}, \textit{citing patents} and \textit{scientific citations} and to averages for the other columns.
We double-count patents that fall into multiple adaptation technology categories, i.e. totals in row \emph{All} are smaller than the sum of totals by technology type. The rows \textit{Gov. support: Yes} and \textit{No} show statistics for the subset of patents that do and do not rely on public support, respectively. The data on government support is only available for the period 1928-2017, i.e. the patent counts do not sum up. 

\end{table}

Table \ref{tab:overview_mitigation_patents} summarizes all Y02-tagged technologies in our data differentiating between different types of mitigation and adaptation technologies. Fig. \ref{fig:family_based_pie_chart_all} shows a pie-chart illustrating the relative frequencies of different types of mitigation and adaptation technologies at the aggregate and disaggregate level.
\begin{table}[h]
{\centering
\begingroup\scriptsize
\caption{Overview of adaptation and mitigation technologies}
\label{tab:overview_mitigation_patents}
\begin{tabular}{p{2.5cm}p{1.45cm}p{1.45cm}p{1.45cm}p{1.45cm}p{1.45cm}p{1.45cm}p{1.45cm}}
\hline
Technology & Total patents & Citing patents & Citing/total patents & Scientific citations & Citations per patents (all) & Citations per patents (citing) & Share Gov. support \\ 
  \hline
Adaptation & 37341 & 14681 & 0.39 & 672420 & 18.01 & 45.80 & 0.31 \\ 
  Buildings & 43371 & 6243 & 0.14 & 39477 & 0.91 & 6.32 & 0.16 \\ 
  CCS & 5111 & 2074 & 0.41 & 22261 & 4.36 & 10.73 & 0.34 \\ 
  Green ICT & 32735 & 8281 & 0.25 & 55787 & 1.70 & 6.74 & 0.11 \\ 
  Energy & 166061 & 40075 & 0.24 & 417128 & 2.51 & 10.41 & 0.32 \\ 
  Production & 83967 & 27315 & 0.33 & 310779 & 3.70 & 11.38 & 0.21 \\ 
  Transport & 109997 & 10457 & 0.10 & 70373 & 0.64 & 6.73 & 0.19 \\ 
  Waste & 22368 & 4361 & 0.19 & 33671 & 1.51 & 7.72 & 0.19 \\ 
   \hline
All & 445689 & 98084 & 0.26 & 1621896 & 4.17 & 13.23 & 0.23 \\ 
\hline
Gov. support (No) & 272606 & 46150 & 0.17 & 378647 & 1.39 & 8.20 &  \\ 
  Gov. support (Yes) & 81245 & 36919 & 0.45 & 1005000 & 12.37 & 27.22 &  \\ 
   \hline
\end{tabular}
\endgroup
}

\footnotesize
Notes: This table summarizes the characteristics of adaptation and mitigation technologies at the aggregate 4-digit CPC level. 
The column \textit{Citing patents} shows the number of patents that rely on science, i.e. make at least one citation to the scientific literature.  
The row entry \textit{All} corresponds to total numbers for columns \textit{total patents}, \textit{citing patents} and \textit{scientific citations} and to averages for the other columns.
We double-count patents that fall into multiple technology categories, i.e. totals in row \emph{All} are smaller than the sum of totals by technology type. 
The rows \textit{Gov. support: Yes} and \textit{No} show statistics for the subset of patents that do and do not rely on public support, respectively.
\end{table}

\FloatBarrier
\subsection{The scientificness of adaptation and mitigation over time}
\label{SI:scientificness}
\begin{figure}[h]
    {\centering
        \caption{Adaptation patents and their science reliance}
    \label{fig:ts_patents_counts_science_reliant}

    \includegraphics[width=\textwidth]{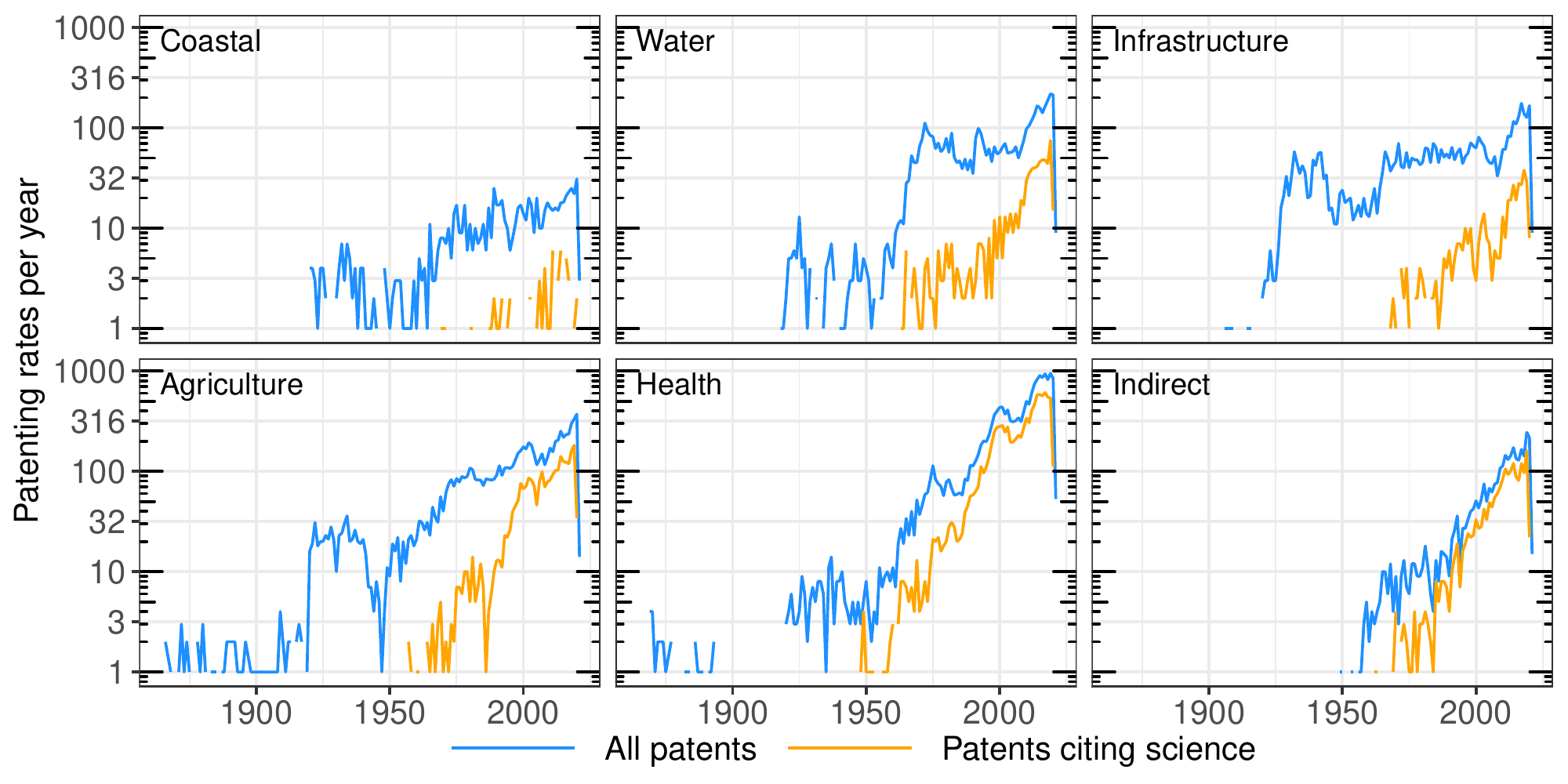}

    }
    \footnotesize
    Notes: Total number of patents by technology type and number of these patents that cite to science over time at a logarithmic scale. 

\end{figure}

\begin{figure}[h]
    {\centering
        \caption{Adaptation and mitigation patents and their science reliance}
    \label{fig:ts_patents_counts_science_reliant_all}

    \includegraphics[width=\textwidth]{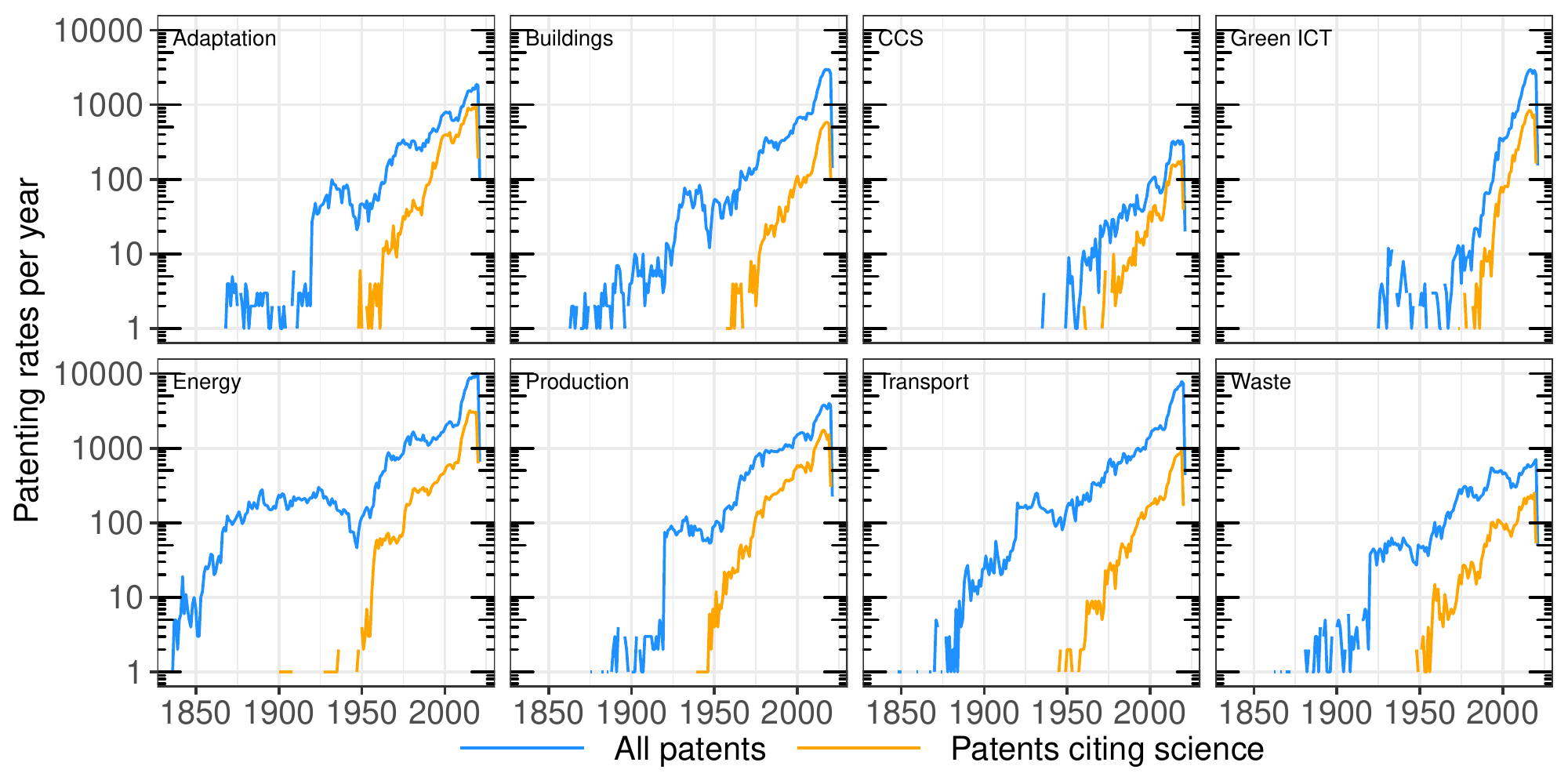}

    }
    \footnotesize
    Notes: Total number of patents by technology type and number of these patents that cite to science over time at a logarithmic scale. 

\end{figure}

Fig. \ref{fig:ts_patents_counts_science_reliant} (Fig. \ref{fig:ts_patents_counts_science_reliant_all}) shows time series plots of counts of 6-digit adaptation (4-digit adaptation and mitigation) patents (blue line) and counts of patents that cite at least one scientific paper (orange line) at a logarithmic scale. 

Among the adaptation technologies, adaptation in agriculture and health are the oldest technologies with first patents being granted in the mid 19th century. Indirect adaptation technologies emerged in the 1960s and exhibit a strong reliance on science. While agriculture, health and indirect exhibit exponential growth, the other three technologies rather stagnated for a long time. Post-2005, the number of annually granted patents was increasing. 

Clean energy technologies show historically a relatively high number of patents, and also for adaptation, mitigation related to buildings, production and transport, patenting began already during the nineteenth century. Green ICT and CCS are the by far youngest technologies starting off in early to mid-twentieth century. 
For all technologies, we observe an increasing reliance on science starting off from the 1950s. The reliance on science has been increasing for all technologies, though we observe a strong heterogeneity across technology groups with adaptation, clean production, and CCS having the highest share of patents that make at least once citation to a scientific article (see also Table \ref{tab:overview_mitigation_patents}). 

\begin{figure}[h]
    {\centering
        \caption{Science intensity of adaptation and mitigation}
    \label{fig:ts_scientificness_tech_all}

    \includegraphics[width=\textwidth]{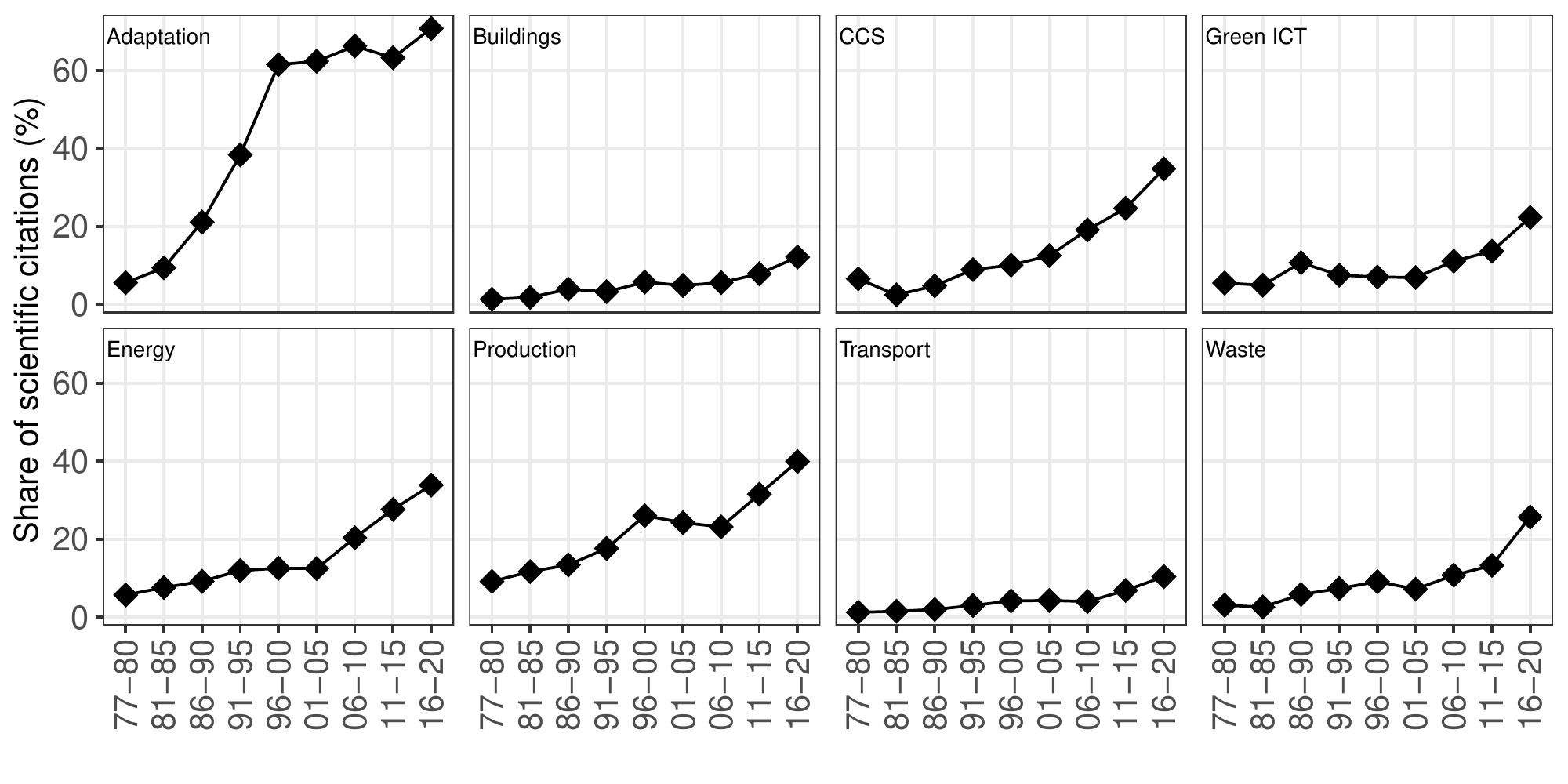}

    }
    \scriptsize
    Notes: Science-intensity of different 4-digit adaptation and mitigation patents measured by the share of citations to science in the number of total citations (sum of citations to other patents and scientific articles). 

\end{figure}

In Fig. \ref{fig:ts_scientificness_tech_all} we show an alternative measure of the scientificness of patents given by the ratio of citations to science over the sum of citations to patents and science. This figure confirms the pattern observed before with adaptation showing the highest reliance on scientific rather than applied knowledge, but also CCS, clean production, energy, waste and green ICT to become increasingly scientific. However, as seen in Section \ref{sec:science_base}, the there is a high heterogeneity across subfields as technology. For example, the high science intensity of adaptation is mainly driven by health technologies and previous research has show that solar PV and biofuels are key drivers of the scientificness of clean energy technologies \citep{hotte2021rise}.

\FloatBarrier


\newpage
\renewcommand{\appendixname}{Online Supplementary Material}
\renewcommand{\thesection}{SI.\arabic{section}} \setcounter{section}{0}
\renewcommand{\thefigure}{SI.\arabic{figure}} \setcounter{figure}{0}
\renewcommand{\thetable}{SI.\arabic{table}} \setcounter{table}{0}
\renewcommand{\theequation}{SI.\arabic{table}} \setcounter{equation}{0}

\FloatBarrier 

\bigskip
\begin{center}
    \LARGE 
    \vspace{8cm}
    Online Supplementary Material\\
    \vspace{2cm} \normalsize
    
   \large
    \emph{Knowledge for a warmer world: A patent analysis of climate change adaptation technologies}\\ \vspace{1cm}
    \normalsize
    Kerstin H\"otte, Su Jung Jee, Sugandha Srivastav
\end{center}

\newpage

\section{Relative frequencies of adaptation and mitigation patents}

\begin{figure}[H]
{    \centering
    \includegraphics[width=\textwidth]{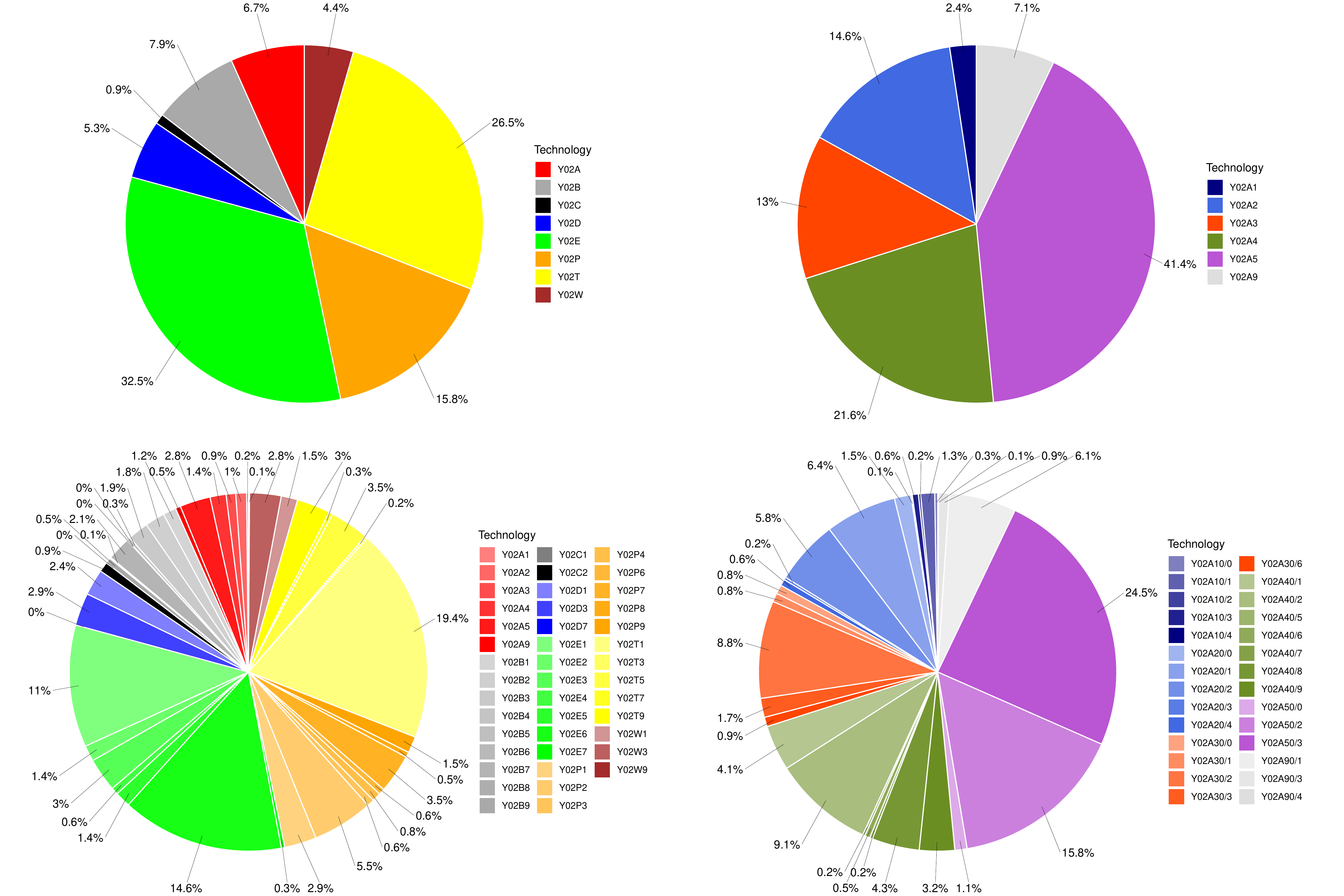}
    \caption{Relative frequencies of different technologies}
    \label{fig:family_based_pie_chart_all}
    }
    \footnotesize
    Notes: These pie charts illustrate the relative frequencies of different types of mitigation and adaptation technologies. The figures at the left show pie charts for all Y02-tagged technologies (with adaptation technologies (Y02A) indicated in red color). The figures at the right show these numbers for the subset of adaptation technologies. 
\end{figure}

\FloatBarrier
\subsection{Additional information on the science base of adaptation}
\label{SI:add_info_science_base}
Table \ref{tab:overview_science_citations} gives an overview of the characteristics of the data on citation links from patents to science in our data showing the number of patents that cite science, the number of different papers being cited, the number of citation links, the number of citation links with the highest reliability score, the average confidence score, the percentage share of citations made by the applicant and by the examiner, and the percentage of citations made on the front page of the patent and in the text body. Interestingly, science citations in science-intensive technologies (agriculture, health, indirect) are most often made by the applicant (63.6-83.5\%) while citation links in technologies with low science reliance (coastal, water, infrastructure) are mostly added by the patent examiner (24.4-35.7\% of citations made by the applicant).  
\begin{landscape}
\begin{table}[ht]
{\centering
\caption{Overview statistics of the subset of science-reliant adaptation patents and their citation links}
\label{tab:overview_science_citations}
\begingroup\footnotesize
\begin{tabular}{p{3cm}p{1.75cm}p{1.75cm}p{1.75cm}p{1.75cm}p{1.75cm}p{1.75cm}p{1.75cm}p{1.75cm}p{1.75cm}}
\hline
Technology & \# patents & \# papers & \# citations & \# CS = 10 & Avg CS & \% app & \% exm & \% text & \% front \\ 
  \hline
\hline
Coastal &  62 & 225 & 328 & 284 & 9.69 & 35.67 & 64.33 & 15.55 & 81.10 \\ 
   \hline
Gov. support (No) &  31 &  70 & 115 &  88 & 9.37 & 25.22 & 74.78 & 17.39 & 81.74 \\ 
  Gov. support (Yes) &  28 & 154 & 204 & 187 & 9.85 & 40.69 & 59.31 & 12.75 & 82.35 \\ 
   \hline
\hline
Water & 720 & 3967 & 14173 & 13609 & 9.91 & 32.46 & 67.54 & 10.10 & 77.50 \\ 
   \hline
Gov. support (No) & 299 & 858 & 2346 & 1967 & 9.61 & 36.91 & 63.09 & 19.82 & 68.12 \\ 
  Gov. support (Yes) & 287 & 2405 & 8493 & 8344 & 9.97 & 29.59 & 70.41 & 9.22 & 85.46 \\ 
   \hline
\hline
Infrastructure & 421 & 1651 & 3828 & 3648 & 9.90 & 24.40 & 75.60 & 11.65 & 83.10 \\ 
   \hline
Gov. support (No) & 171 & 405 & 1108 & 1004 & 9.82 & 29.78 & 70.22 & 18.23 & 74.64 \\ 
  Gov. support (Yes) & 175 & 836 & 2014 & 1967 & 9.94 & 20.56 & 79.44 & 7.75 & 88.53 \\ 
   \hline
\hline
Agriculture & 2482 & 40962 & 149341 & 149039 & 9.99 & 83.46 & 16.39 & 66.52 & 23.30 \\ 
   \hline
Gov. support (No) & 1114 & 15873 & 51788 & 51596 & 9.99 & 91.21 & 8.79 & 80.91 & 13.16 \\ 
  Gov. support (Yes) & 991 & 27196 & 77459 & 77377 & 10.00 & 80.00 & 19.71 & 57.29 & 29.28 \\ 
   \hline
\hline
Health & 9331 & 238282 & 490310 & 489397 & 10.00 & 67.59 & 32.36 & 36.20 & 46.39 \\ 
   \hline
Gov. support (No) & 3638 & 62045 & 94135 & 93488 & 9.99 & 78.22 & 21.78 & 55.45 & 30.14 \\ 
  Gov. support (Yes) & 4495 & 170407 & 336138 & 335912 & 10.00 & 65.33 & 34.59 & 31.38 & 50.31 \\ 
   \hline
\hline
Indirect & 1749 & 44865 & 72759 & 72596 & 10.00 & 63.58 & 36.42 & 30.75 & 55.82 \\ 
   \hline
Gov. support (No) & 495 & 7666 & 9405 & 9333 & 9.98 & 82.74 & 17.26 & 64.01 & 26.87 \\ 
  Gov. support (Yes) & 974 & 34299 & 56303 & 56240 & 10.00 & 63.29 & 36.71 & 26.25 & 59.70 \\ 
   \hline
\hline
All & 14681 & 313244 & 730739 & 728573 & 9.99 & 69.51 & 30.43 & 41.21 & 43.42 \\ 
   \hline
Gov. support (No) & 5717 & 84017 & 158897 & 157476 & 9.98 & 81.73 & 18.26 & 63.44 & 25.32 \\ 
  Gov. support (Yes) & 6911 & 223825 & 480611 & 480027 & 10.00 & 66.63 & 33.28 & 34.46 & 48.82 \\ 
   \hline
\end{tabular}
\endgroup

}

\footnotesize
Notes: Columns show the number of (1) unique patents, (2) unique papers, (3) citation links, (4) citations with highest confidence score (CS = 10), (5) the average confidence score, (6) the share of applicant and (7) examiner added citations (remaining share is of unknown type), (8) share of citations made exclusively in the text body or (9) front page of the patent document (remaining shares account for citations made in both).
\end{table}
\end{landscape}

Table \ref{tab:citing_patents_age_adapt} summarizes the age characteristics of science-reliant patents and papers being cited. The oldest science reliant patents were granted between 1935 (agriculture) and 1965 (coastal). The oldest scientific papers cited are from the nineteenth century. 
\begin{table}[ht]
{
\centering
\caption{Age characteristics of science-reliant adaptation patents and cited papers}
\label{tab:citing_patents_age_adapt}
\begingroup\footnotesize
\begin{tabular}{p{2.75cm}p{1.25cm}p{1.25cm}p{1.5cm}|p{1.25cm}p{1.25cm}p{1.5cm}p{1.25cm}}
\hline
\multicolumn{1}{p{1.25cm}}{Technology}&\multicolumn{1}{p{1.25cm}}{Oldest patent}&\multicolumn{1}{p{1.25cm}}{Youngest patent}&\multicolumn{1}{p{1.5cm}}{Avg year patent}&\multicolumn{1}{p{1.25cm}}{Oldest paper}&\multicolumn{1}{p{1.25cm}}{Youngest paper}&\multicolumn{1}{p{1.5cm}}{Avg year paper}&\multicolumn{1}{p{1.25cm}}{Avg lag}
\\
\hline
Coastal & 1965 & 2020 & 2006.04 & 1856 & 2015 & 1993.93 & 12.11 \\ 
Water & 1963 & 2020 & 2012.39 & 1867 & 2018 & 1998.38 & 14.01 \\ 
Infrastructure & 1939 & 2020 & 2011.29 & 1931 & 2018 & 1999.21 & 12.07 \\ 
Agriculture & 1935 & 2020 & 2009.84 & 1855 & 2018 & 1993.06 & 16.77 \\ 
Health & 1948 & 2020 & 2010.02 & 1831 & 2019 & 1995.48 & 14.54 \\ 
Indirect & 1949 & 2020 & 2010.57 & 1879 & 2019 & 1996.91 & 13.66 \\ 
\hline
All & 1935 & 2020 & 2010.09 & 1831 & 2019 & 1995.20 & 14.88 \\ 
   \hline
\end{tabular}
\endgroup

}

\footnotesize
Notes: This table summarizes information on the age of science-reliant patents and cited papers. Columns show the (1) grant year of the oldest patent, (2) grant year of the most recent patent, (3) average grant year of patents, (4) publication year of the oldest cited paper, (5) publication year of the most recent cited paper, (6) average publication year of papers and (7) average citation lag.
\end{table}

Table \ref{tab:top_WoS_adapt} the most frequently cited scientific fields by adaptation technology. Biochemistry and molecular biology turn out to be the by far most often cited field of research. 
\begin{table}[ht]
{
\centering
\caption{Most frequently cited WoS fields}
\label{tab:top_WoS_adapt}
\begingroup\footnotesize
\begin{tabular}{p{2.75cm}p{5cm}p{1.75cm}p{1.75cm}}
\hline
Technology & WoS field & \# citations (by tech) & \# citations (total) \\ 
\hline
Coastal & Environmental Sciences &  39 & 3,660 \\ 
Water & Eng., Chem. & 2,777 & 4,368 \\ 
Infrastructure & Energy \& Fuels & 474 & 3,000 \\ 
Agriculture & Biochem. \& Molec. Biology & 68,260 & 231,750 \\ 
Health & Biochem. \& Molec. Biology & 144,666 & 231,750 \\ 
Indirect & Biochem. \& Molec. Biology & 18,447 & 231,750 \\ 
\hline
All & Biochem. \& Molec. Biology & 231,750 & 231,750 \\ 
\hline
\end{tabular}
\endgroup

}
\footnotesize

Notes: This table shows the number of citations from patents to different scientific fields. The third and fourth columns show the number of citations that the WoS field has received by patents in each adaptation technology field and by patent from all adaptation technology fields, respectively.
\end{table}

Table \ref{tab:top_journals_adapt} shows the most important journals cited by adaptation patents. Table \ref{tab:top_patent_adapt} shows the most science-reliant patent, i.e. the patent making most scientific citations and table \ref{tab:top_papers_adapt} shows the papers that have been most important for different types of adaptation technologies. 
\begin{table}[h]
{\centering
\caption{Most frequently cited journals}
\label{tab:top_journals_adapt}
\begingroup\footnotesize
\begin{tabular}{p{2cm}p{3.5cm}p{1.75cm}p{1.75cm}}
\hline
Technology & Journal  & \# citations (by tech) & \# citations (total) \\ 
  \hline
Coastal & Electrochimica Acta &  10 &  10 \\ 
Water & Desalination & 1,087 & 1,087 \\ 
Infrastructure & Science & 113 & 113 \\ 
Agriculture & PNAS & 10,483 & 10,483 \\ 
Health & PNAS & 22,732 & 22,732 \\ 
Indirect & PNAS & 2,967 & 2,967 \\ 
\hline
All & PNAS & 36,267 & 36,267 \\ 
\hline
\end{tabular}
\endgroup

}

\footnotesize
Notes: This table shows the journals that are cited most frequently. 
The third and fourth columns show the number of citations that the journal has received by patents in each adaptation technology field and in all adaptation technology fields, respectively. PNAS is the abbreviation of Proceedings of the National Academy of Sciences of the United States of America.
\end{table}

\begin{table}[ht]
{
\centering
\caption{Most frequently cited papers}
\label{tab:top_papers_adapt}
\begingroup\scriptsize
\begin{tabular}{p{2cm}p{6cm}p{1cm}p{2cm}p{1.125cm}p{1.125cm}}
\hline
Technology & Paper title & Year & WoS field & \# cit. (techn) & \# cit. (total) \\ 
  \hline
\hline
Coastal & Characterization Of Dredged River Sediments In 10 Upland Disposal Sites Of Alabama & 1995 & Construction \& Building Tech. &  5 &   5 \\ 
Water & The Philips Stirling Engine & 1991 & Thermodynamics &  66 &  66 \\ 
Infrastructure & Conventional Wallboard With Latent Heat Storage For Passive Solar Applications & 1990 & Thermodynamics &  23 &  23 \\ 
Agriculture & Identification Of Dna Sequences Required For Activity Of The Cauliflower Mosaic Virus 35s Promoter & 1985 & Plant Sciences &  484 & 505 \\ 
Health & Antibodies A Laboratory Manual & 1988 & Computer Sci., Artif. Intell. & 547 & 813 \\ 
Indirect & Antibodies A Laboratory Manual & 1988 & Computer Sci., Artif. Intell. & 61 & 813 \\ 
\hline
All & Basic Local Alignment Search Tool & 1990 & Biochem. \& Molec. Biology & 908 & 908 \\ 
\hline
\end{tabular}
\endgroup

}

\footnotesize
Notes: This table shows the papers that are most frequently cited by adaptation patents. Year indicates the publication year of the paper and WoS field indicates the Web-of-Science field of research by which the paper is categorized. The fifth and sixth columns show the number of citations that the paper has received by patents in each adaptation technology field and in all adaptation technology fields, respectively.

\end{table}

\begin{table}[ht]
{\centering
\caption{Patents with highest number of citations to papers}
\label{tab:top_patent_adapt}
\begingroup\scriptsize
\begin{tabular}{p{2cm}p{1.25cm}p{6.125cm}p{2cm}p{0.875cm}p{0.875cm}}
\hline
Technology & Number & Title & CPC code & Year & \# cit. \\ 
\hline
Coastal & 27361969 & Method For Dewatering Flocculated Materials & Y02A10/00 & 1999 &  35 \\ 
Water & 39942683 & Water Vapor Distillation Apparatus, Method And System & Y02A20/00 & 2011 & 2,486 \\ 
Infrastructure & 35432262 & Insulation Containing Inorganic Fiber And Spherical Additives & Y02A30/00 & 2012 & 286 \\ 
Agriculture & 40456023 & Tomato Line Fir 128-1018 & Y02A40/10 & 2010 & 740 \\ 
Health & 10642047 & Plasmodium Falciparum Thrombospondin-Related Anonymous Proteins (Trap), Fragments And Functional Derivatives & Y02A50/30 & 1998 & 1,614 \\ 
Indirect & 23131391 & Methods Of Determining Polypeptide Structure And Function & Y02A90/10 & 2007 & 3,930 \\ 
\hline
All & 44196417 & Genetic Markers Associated With Drought Tolerance In Maize & Y02A40/132 & 2014 & 3,930 \\ 
   \hline
\end{tabular}
\endgroup

}

\footnotesize
Notes: Columns show (1) patent family number (DOCDB), (2) patent title, (3) full CPC code(s), (4) grant year of the patent and (5) the number of citations made by the patent.
\end{table}

\FloatBarrier
\subsection{Co-classification of adaptation technologies}
\label{SI:techn_convergence}
\begin{figure}[ht]
    {\centering
        \caption{Co-classification of adaptation technologies}
    \label{fig:coclasses_4digit_8}
    \includegraphics[width=\textwidth]{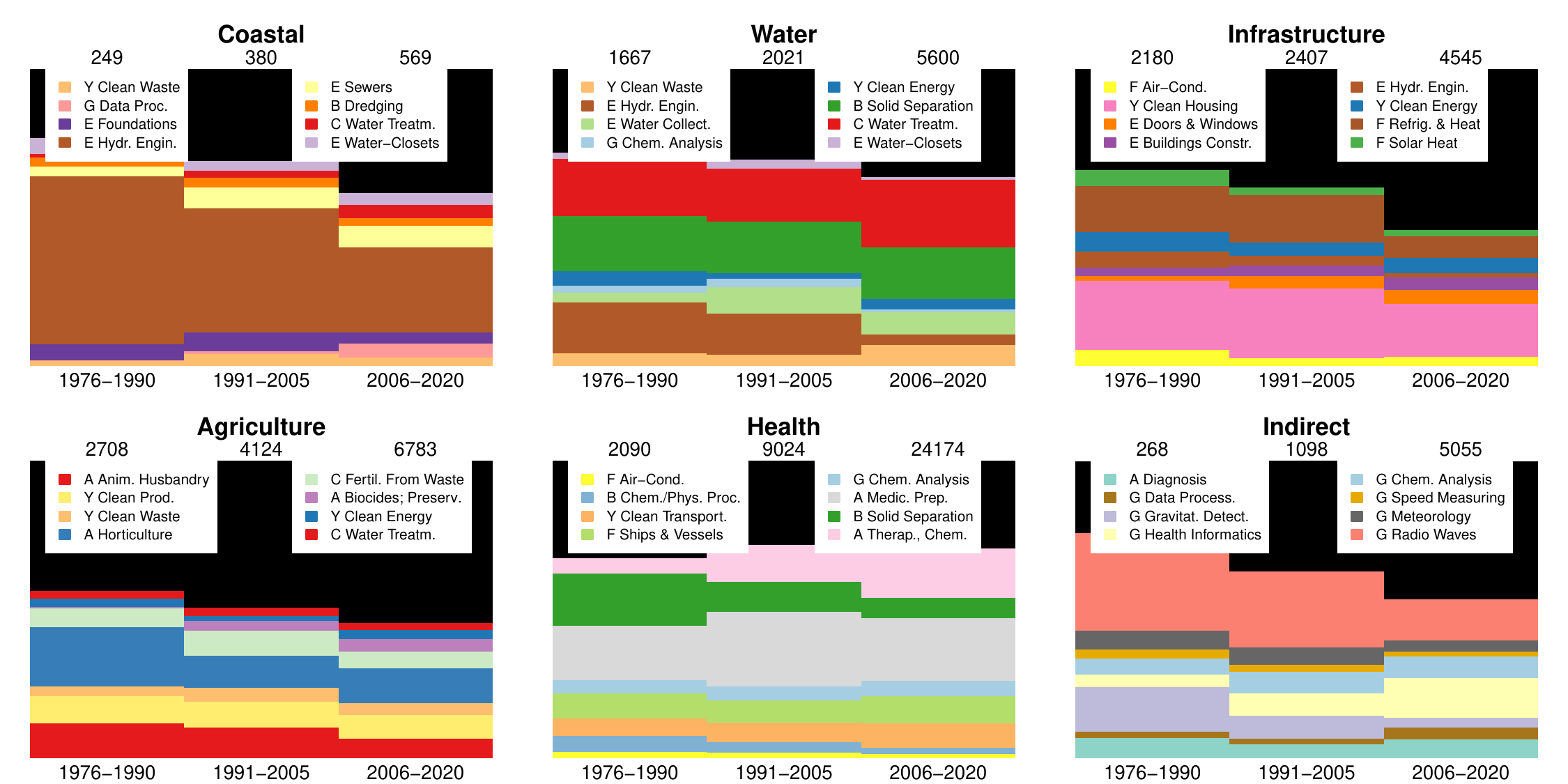}
}

\scriptsize
Notes: These figures show which technology subgroups are most often co-classified with adaptation technologies. Technology subgroups are identified by the 4-digit CPC code. 
The numbers on top of each bar indicate the number of patents granted in each sub-period. Note that the bar plots rely on the number of co-classifications where patents that belong to multiple classes are double-counted. The size of the colored fields in each bar plot indicates the share of co-classifications for different subgroups by adaptation technology type. Co-classifications made for technical reasons (e.g. to ensure the compatibility with other classification systems) are excluded (see \ref{methods:MA_complements}). Black color is used for the residuum of groups that do not belong to the 8 most often co-classified technology groups. The letters in the beginning of the verbal description of each technology group indicate the broad CPC section with A for Human Necessities; B for Performing Operations; C for Chemistry; D for Textiles; E for Fixed Constructions; F for Metallic Engineering; G for Physics and Y as general tagging scheme for cross-sectional technologies (including Y02-tagged green technologies).
\end{figure}

Fig. \ref{fig:coclasses_4digit_8} shows the relative frequencies that adaptation technologies are co-classified with other CPC classes at the 4-digit level. 

\FloatBarrier
\subsection{Dual purpose technologies for mitigation and adaptation}
\label{SI:MA_complements}

\FloatBarrier
\subsubsection{Technological and scientific similarities}
\label{SI:complements_spill}
\begin{figure}
    {\centering
    
        \caption{Scientific \& technological similarities of adaptation \& mitigation in 1976-2005}
    \label{fig:MA_cosine_similarties_plot_76_06}
    \vspace{0.125cm}
        \begin{subfigure}{0.35\textwidth}
            \centering
            \caption{By patent citations (CPC 4-digit)}
            \includegraphics[width=\textwidth]{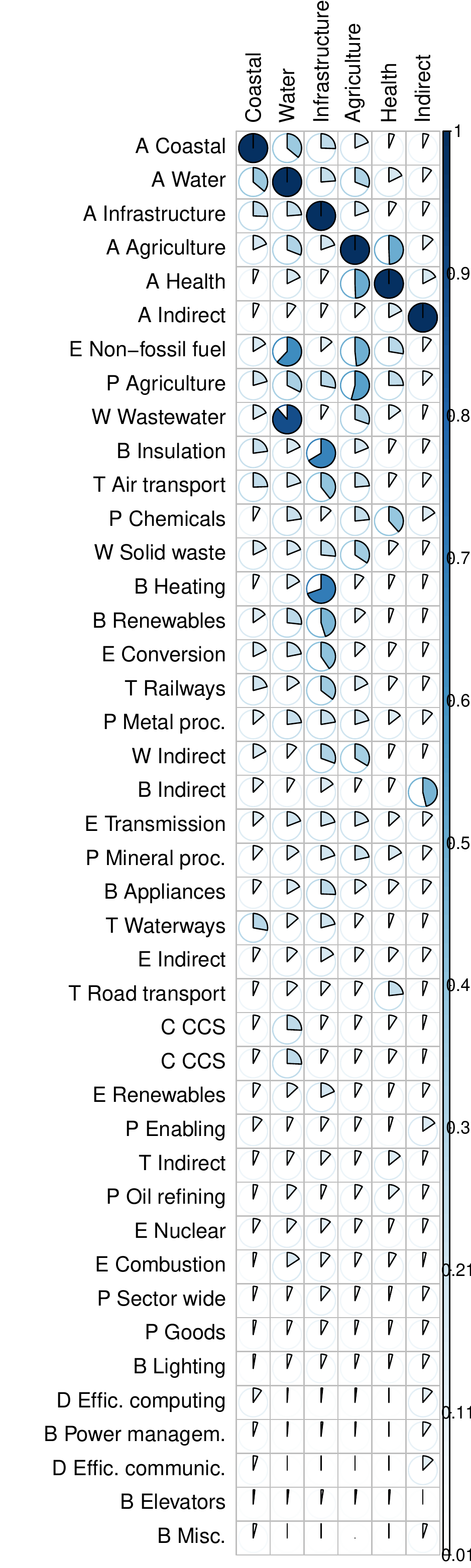}
        \end{subfigure}
        \begin{subfigure}{0.35\textwidth}
            \centering
            \caption{By science citations (WoS fields)}
            \includegraphics[width=\textwidth]{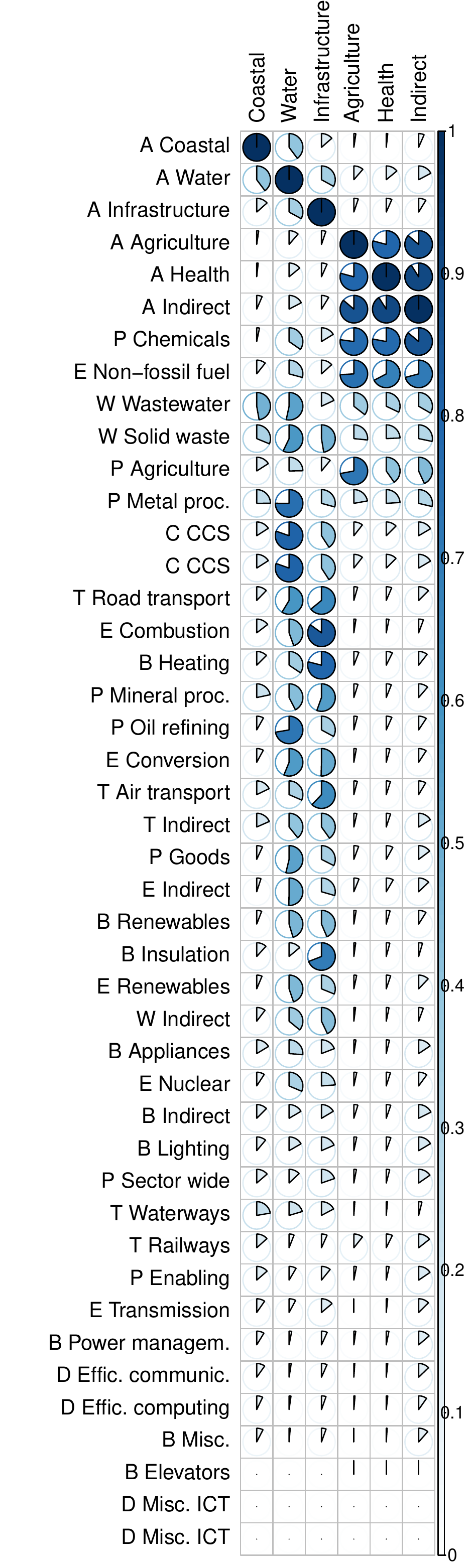}
    \end{subfigure}

    }
    
    \vspace{0.25cm}
    \scriptsize
    Notes: These figures illustrate technological and scientific similarities among adaptation technologies and among adaptation and mitigation technologies at the 6-digit CPC level. We use a cosine similarity-based methodology as explained in \cite{hotte2021rise}. Mitigation technologies are ranked by the row-sum in a decreasing order, i.e. technologies with a higher the spillover potential across all types of adaptation technologies rank higher. Column-wise reading illustrates for which types of adaptation technology the highest spillover potential is found. 
    The letters preceding the name of the technology type indicate the type of 6-digit mitigation or adaptation technology (i.e., A for adaptation, B for buildings, C for capture and storage of greenhouse gases, D for green ICT, E for energy, P for production, T for transport, and W for waste). 
\end{figure}

\begin{figure}
    {\centering
        \caption{Scientific \& technological similarities of adaptation \& mitigation in 2006-2020}
        \vspace{0.125cm}
    \label{fig:MA_cosine_similarties_plot_06_20}
        \begin{subfigure}{0.35\textwidth}
            \centering
            \caption{By patent citations (CPC 4-digit)}
            \includegraphics[width=\textwidth]{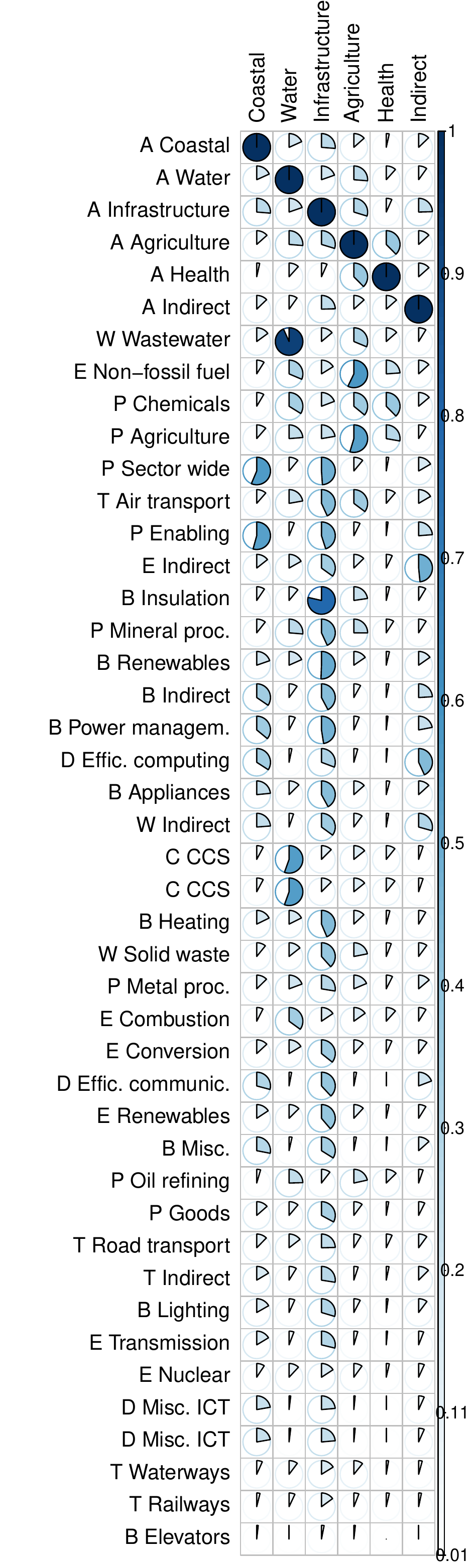}
        \end{subfigure}
        \begin{subfigure}{0.35\textwidth}
            \centering
            \caption{By science citations (WoS fields)}
            \includegraphics[width=\textwidth]{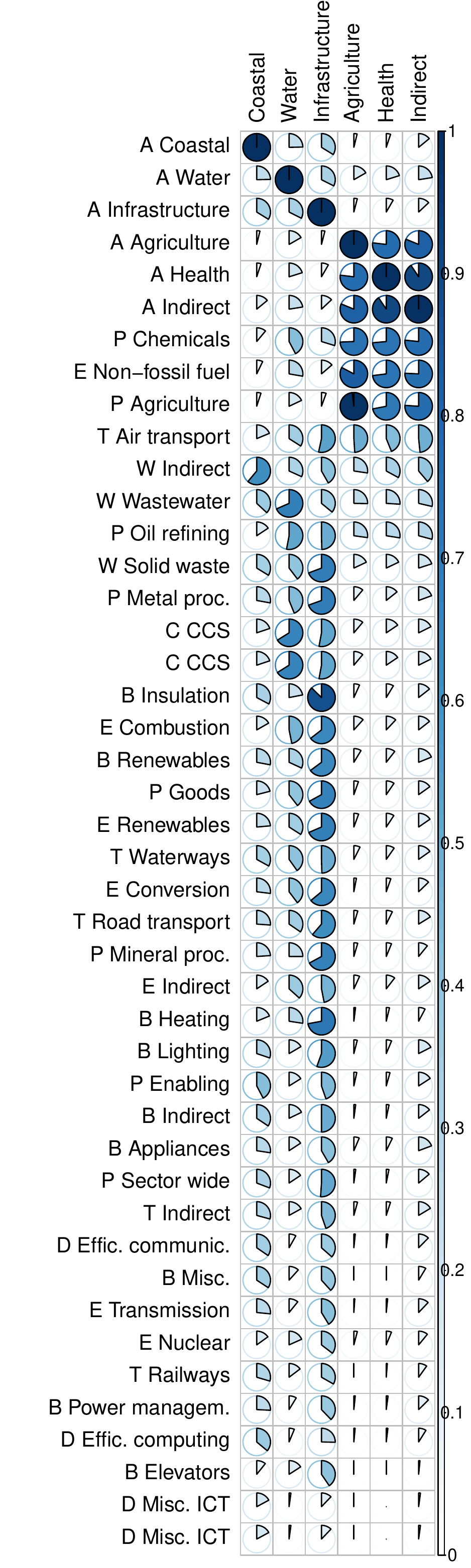}
    \end{subfigure}

    }
    \vspace{0.25cm}
    \scriptsize
    Notes: These figures illustrate technological and scientific similarities among adaptation technologies and among adaptation and mitigation technologies at the 6-digit CPC level. We use a cosine similarity-based methodology as explained in \cite{hotte2021rise}. Mitigation technologies are ranked by the row-sum in a decreasing order, i.e. technologies with a higher the spillover potential across all types of adaptation technologies rank higher. Column-wise reading illustrates for which types of adaptation technology the highest spillover potential is found. 
    The letters preceding the name of the technology type indicate the type of 6-digit mitigation or adaptation technology (i.e., A for adaptation, B for buildings, C for capture and storage of greenhouse gases, D for green ICT, E for energy, P for production, T for transport and W for waste). 
\end{figure}

Technological and scientific knowledge spillovers are likely to occur if knowledge bases of two technologies are sufficiently similar such that the knowledge is mutually useful. 
Fig. \ref{fig:MA_cosine_similarties_plot_76_06} and \ref{fig:MA_cosine_similarties_plot_06_20} show the scientific and technological similarities among adaptation technologies and among adaptation and mitigation technologies at the 6-digit CPC level for the subperiods 1976-2005 and 2006-2020. The rows with mitigation technologies are ordered by their cumulative cosine similarity with adaptation technologies, i.e. by the row sum showing those technologies with the highest overlap with adaptation first. An extract with the fifteen highest scoring technologies is shown in Sec. \ref{sec:mitigation_adaptation} in the main text.

\begin{figure}
{    \centering
    \textbf{Technological similarity by CPC4 citations:} \\
    \includegraphics[width=\textwidth]{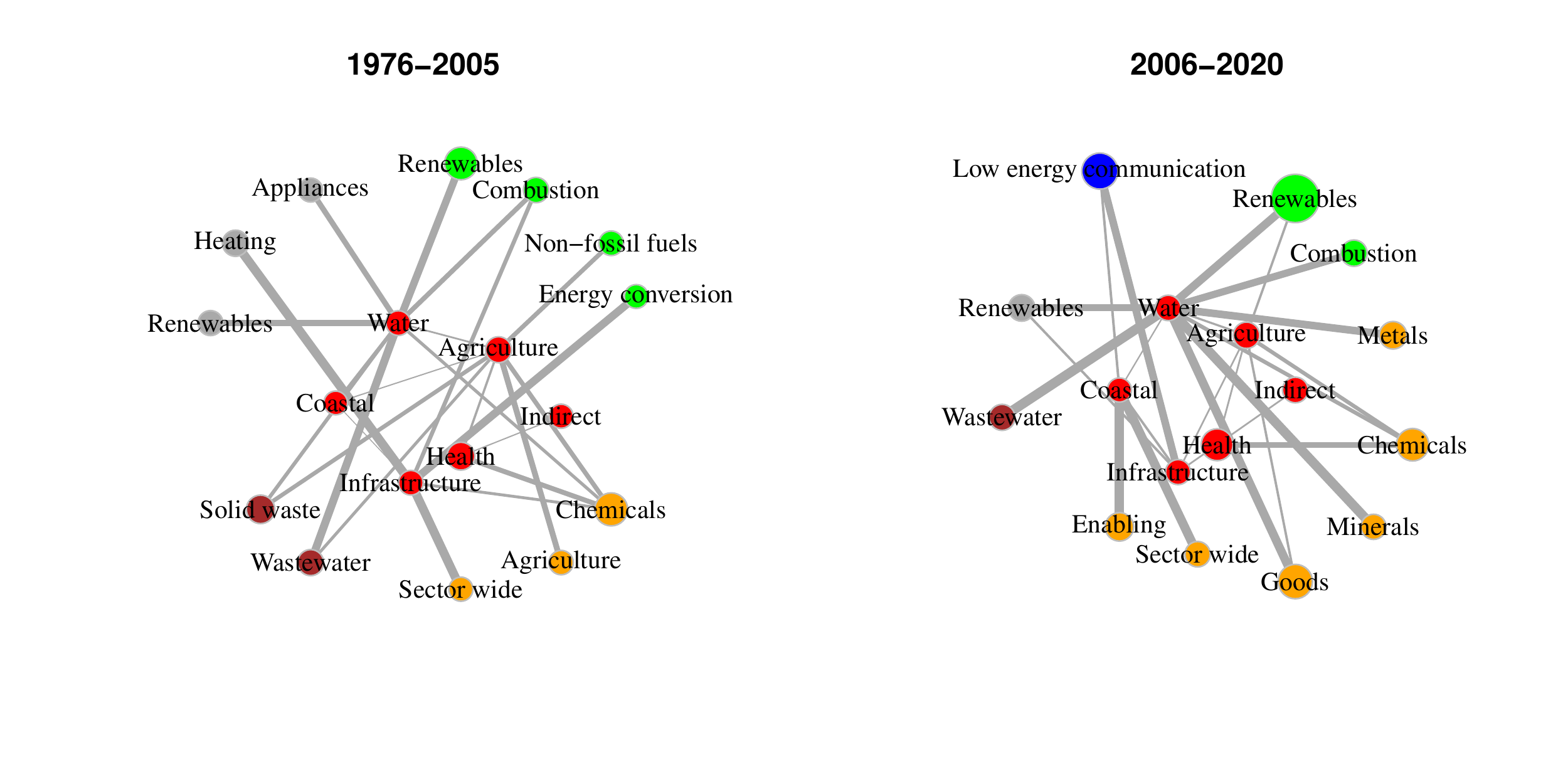}
    
    \textbf{Scientific similarity by WoS field citations:} \\
    \includegraphics[width=\textwidth]{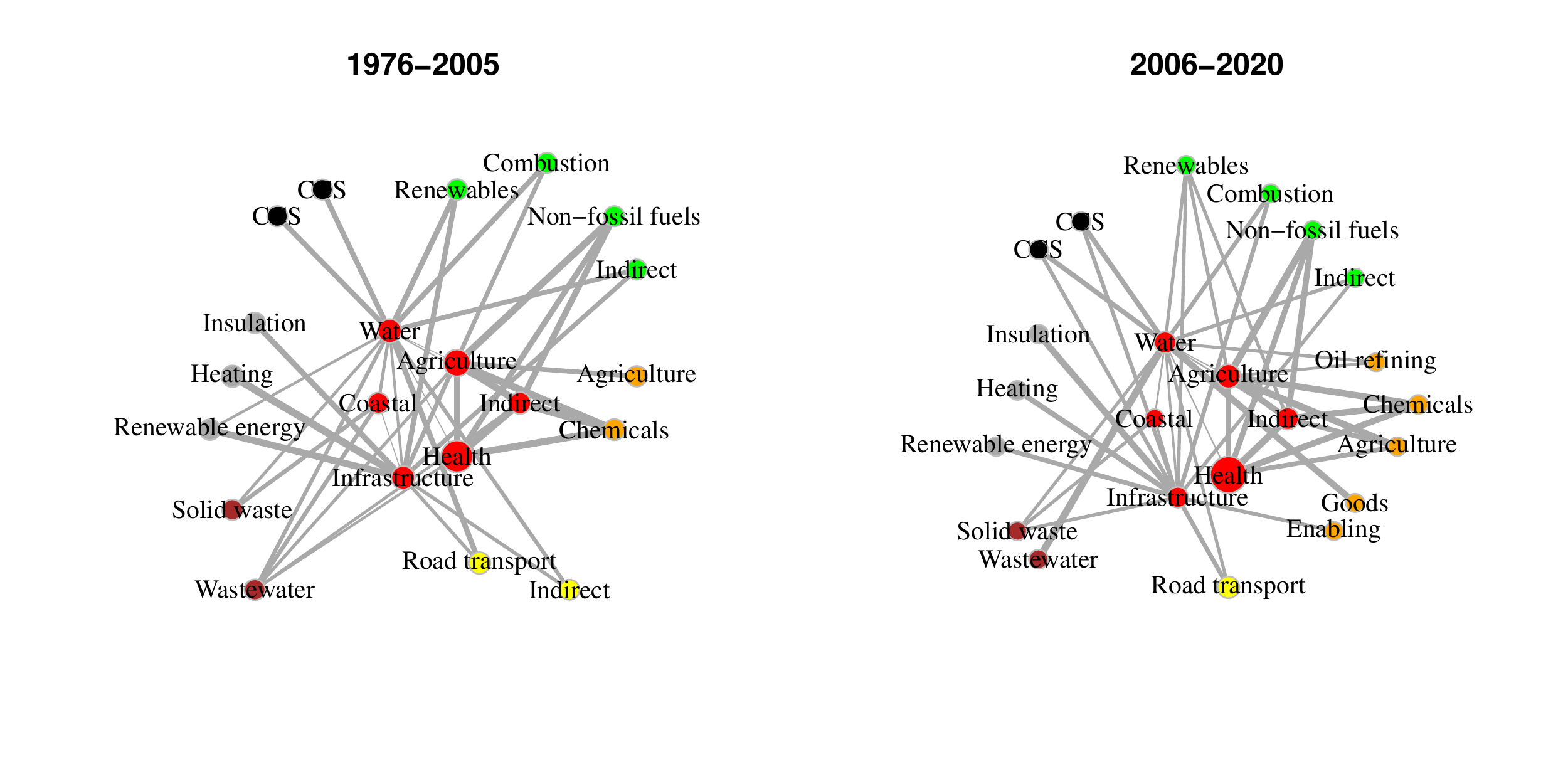}
    
    \caption{Cosine similarity networks of mitigation-adaptation complements}
    \label{fig:family_based_similarity_networks_complements}
    }
    
    \footnotesize 
    Notes: These networks illustrate technological and scientific similarities of different types of adaptation and mitigation technologies. The networks are based on the shares of (a) citations to CPC 4-digit technology classes and (b) citations to scientific fields (WoS). 
    A link between a pair of technologies indicates the cosine similarity of their references to scientific fields (scientific similarity) and technology classes (technological similarity). 
    For clarity only the most significant links are shown. 
    The widths of connecting edges are proportional to the degree of similarity and the node sizes are proportional to the number of patents. 
    The node colors indicate the 4-digit technology class (red for adaptation, gray for buildings, black for CCS, blue for green ICT, green for energy, orange for production, yellow for transport, and brown for waste).
\end{figure}
\begin{figure}
	{    \centering
		\textbf{Technological similarity by CPC4 citations:} \\
		\includegraphics[width=\textwidth]{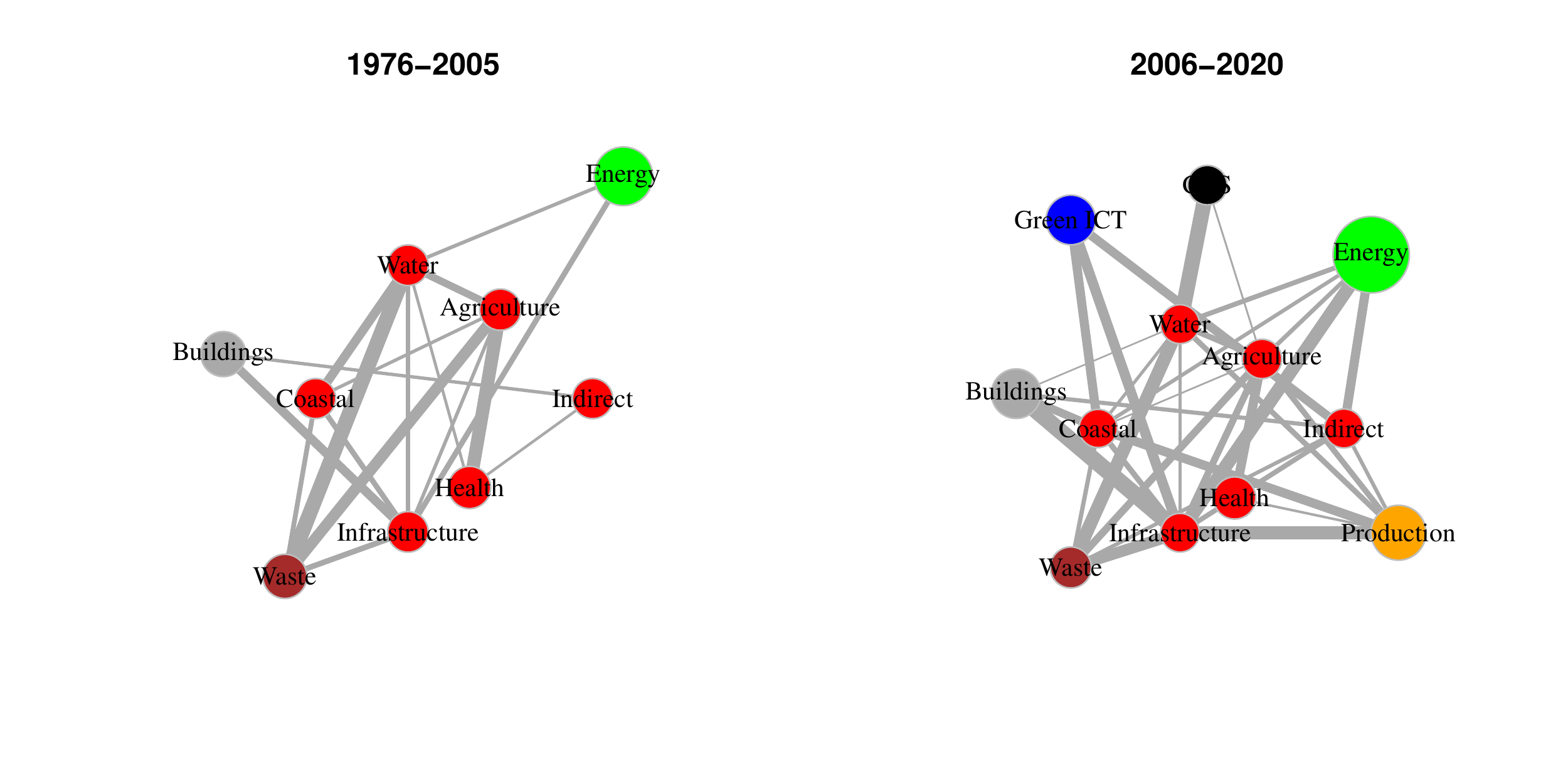}
		
		\textbf{Scientific similarity by WoS field citations:} \\
		\includegraphics[width=\textwidth]{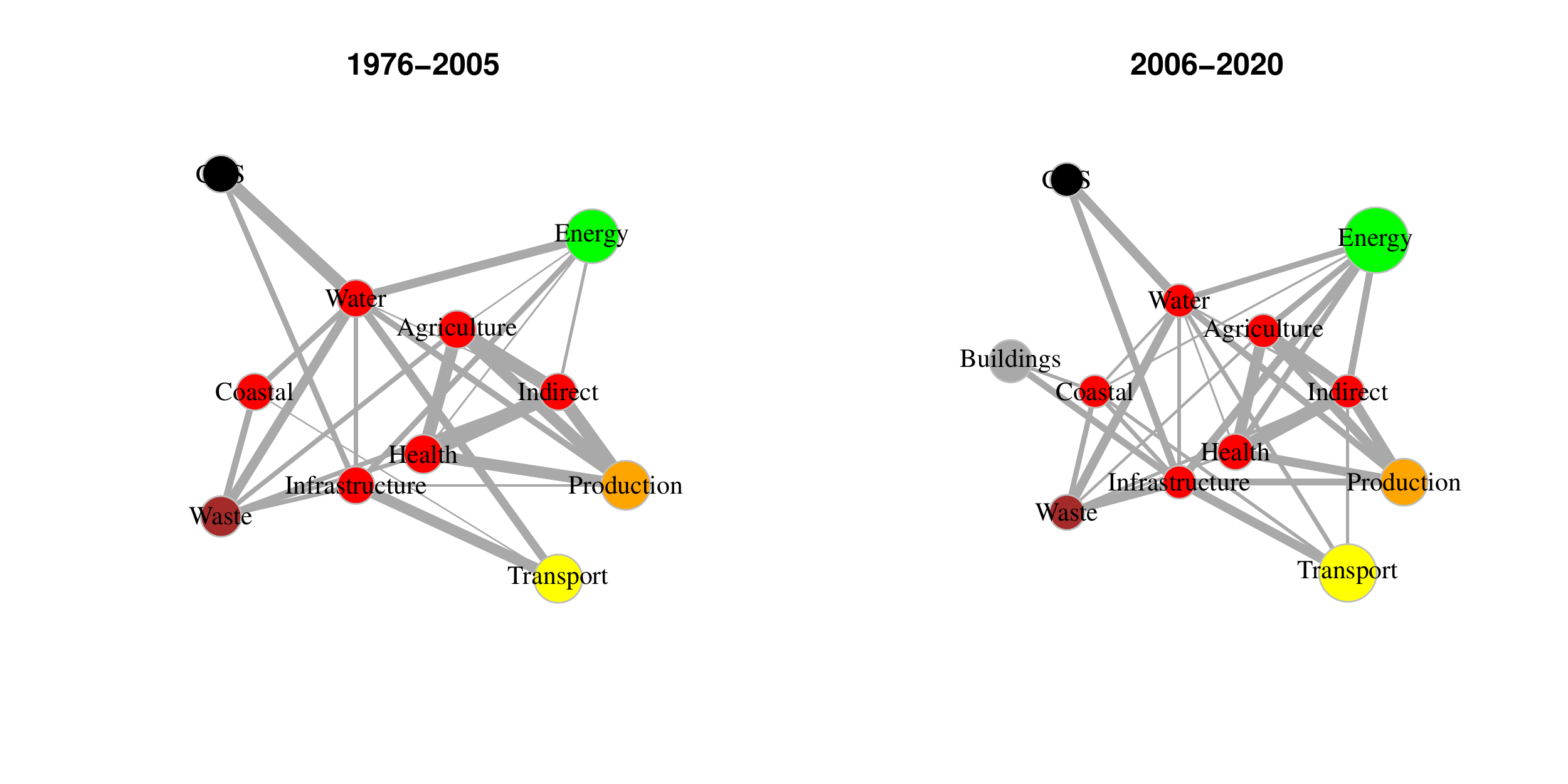}
		
		\caption{Cosine similarity networks of mitigation and adaptation technologies}
		\label{fig:family_based_similarity_networks_ma}
	}
	
	\footnotesize 
	Notes: The networks illustrate the technological and scientific similarity of different adaptation and mitigation technologies computed on the basis of (a) citations to CPC 4-digit technology classes and (b) citations to scientific fields (WoS). A link between a pair of technologies indicates the cosine similarity of their references to scientific fields (scientific similarity) and technology classes (technological similarity). 
	To simplify the representation, only the most significant links are shown. 
	The widths of connecting edges are proportional to the level of similarity and the node sizes are proportional to the number of patents. 
	The node colors indicate the 4-digit technology class (red for adaptation, gray for buildings, black for CCS, blue for green ICT, green for energy, orange for production, yellow for transport and brown for waste). 
\end{figure}

Fig. \ref{fig:family_based_similarity_networks_complements} illustrates scientific and technological similarities between mitigation and adaptation technologies at the disaggregate level (6-digit CPC) using the subset of adaptation patents with mitigation co-benefit. Fig. \ref{fig:family_based_similarity_networks_ma} illustrates technological and scientific similarities of mitigation and adaptation technologies using the full sample of mitigation and adaptation technologies.

\FloatBarrier
\subsubsection{Science base of dual purpose technologies}
\label{SI:MA_complements_science_base}
\begin{figure}[H]
    {\centering
    \includegraphics[width=\textwidth]{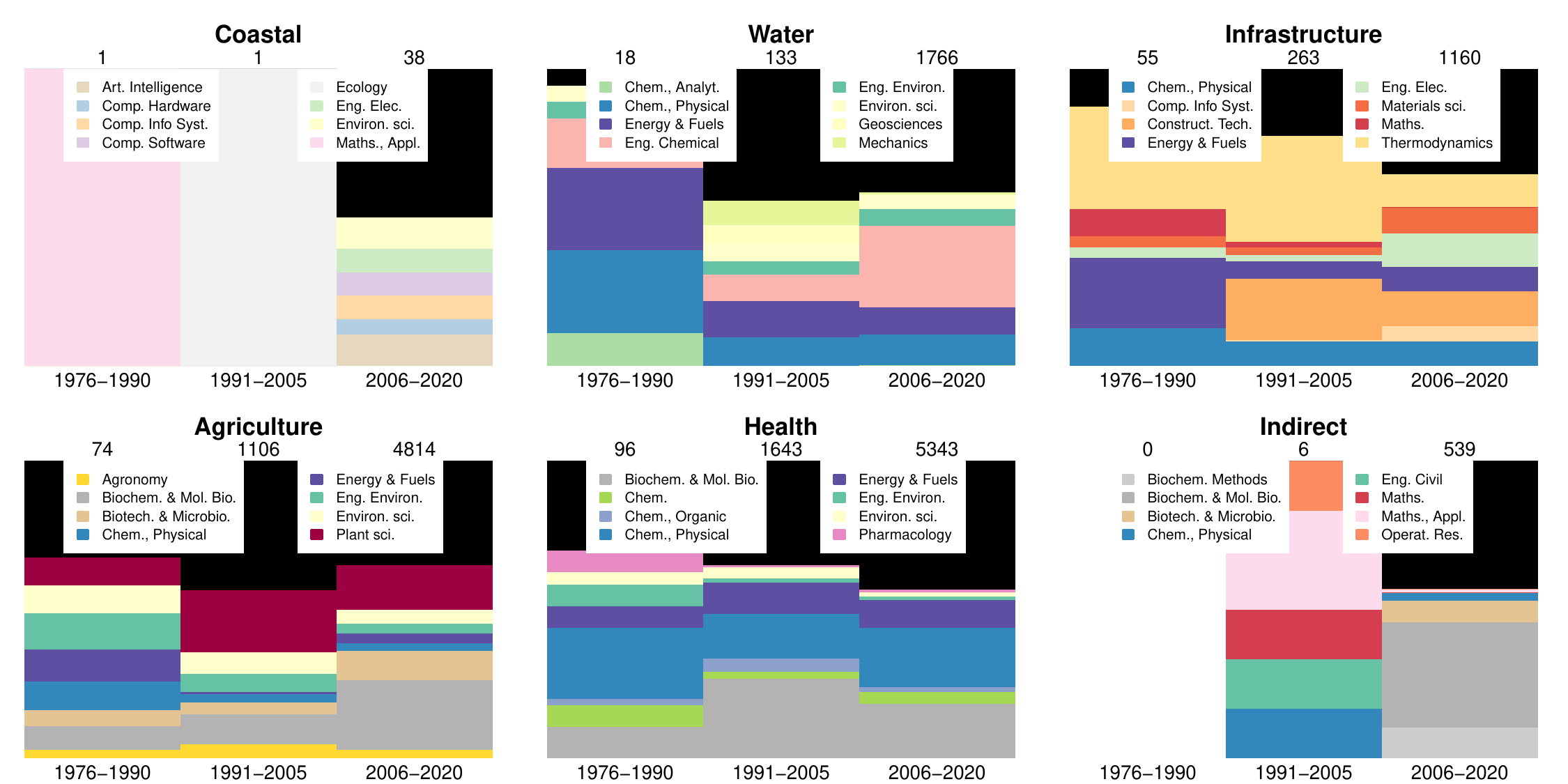}
    \caption{Scientific base of adaptation technologies with mitigation co-benefit}
    \label{fig:SI_science_base_MA_complements_6}
    
    }
    
    \footnotesize
    Notes: This figure shows the 8 scientific fields (Web-of-Science categories) that are most frequently cited by patents for adaptation that have a co-benefit for climate change mitigation.
\end{figure}

\begin{figure}
    {\centering
    \includegraphics[width=\textwidth]{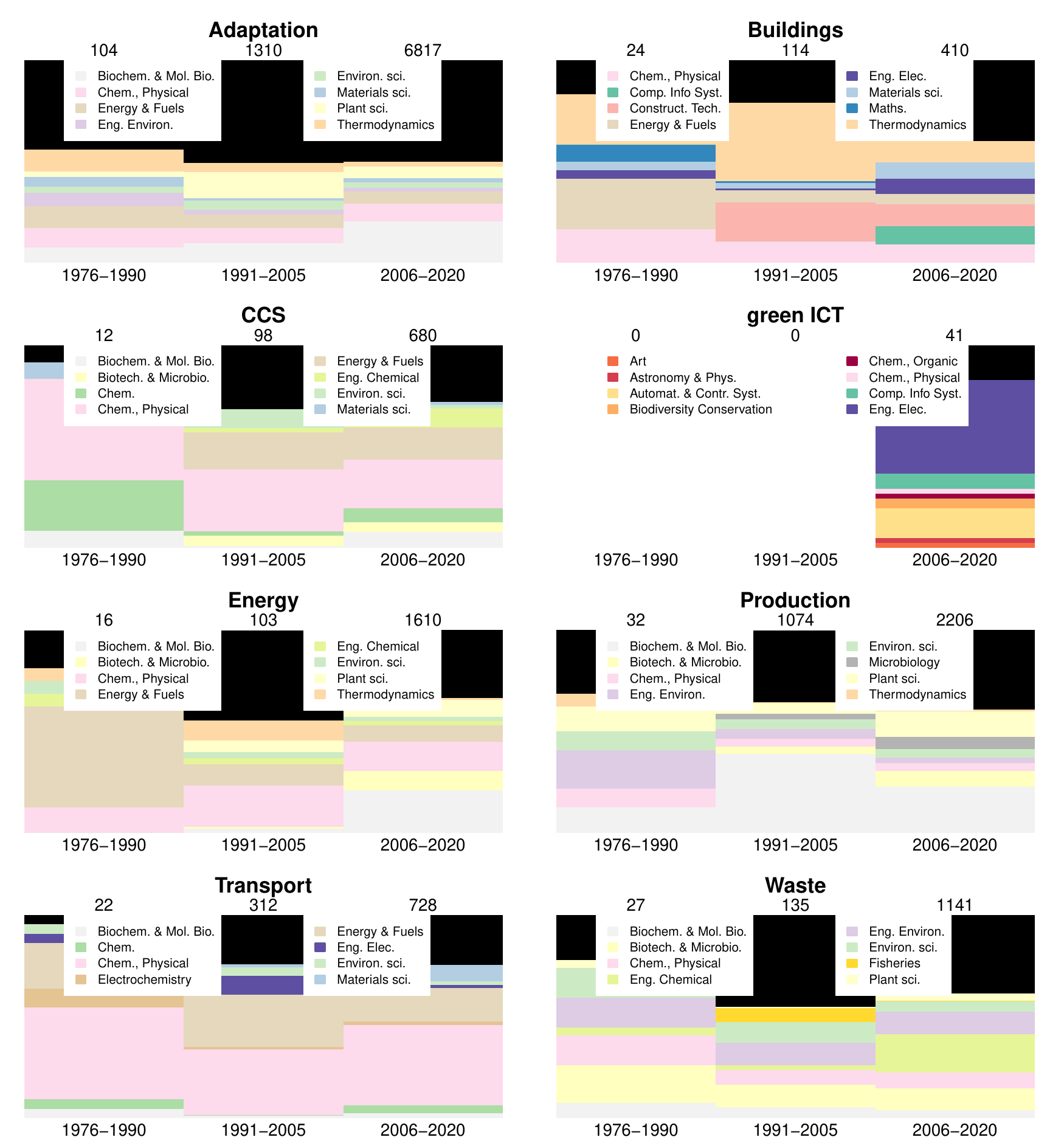}
    \caption{Scientific base of mitigation technologies with adaptation co-benefit}
    \label{fig:SI_science_base_MA_complements_mitigation_6}
    
    }
    
    \footnotesize
    Notes: This figure shows the 8 scientific fields (Web-of-Science categories) that are most frequently cited by patents for mitigation that have a co-benefit for climate change adaptation.
\end{figure}

\FloatBarrier
\subsubsection{Co-classification of dual purpose patents}
\label{SI:MA_complements_techn_convergence}
\begin{figure}[H]
    {\centering
    \includegraphics[width=\textwidth]{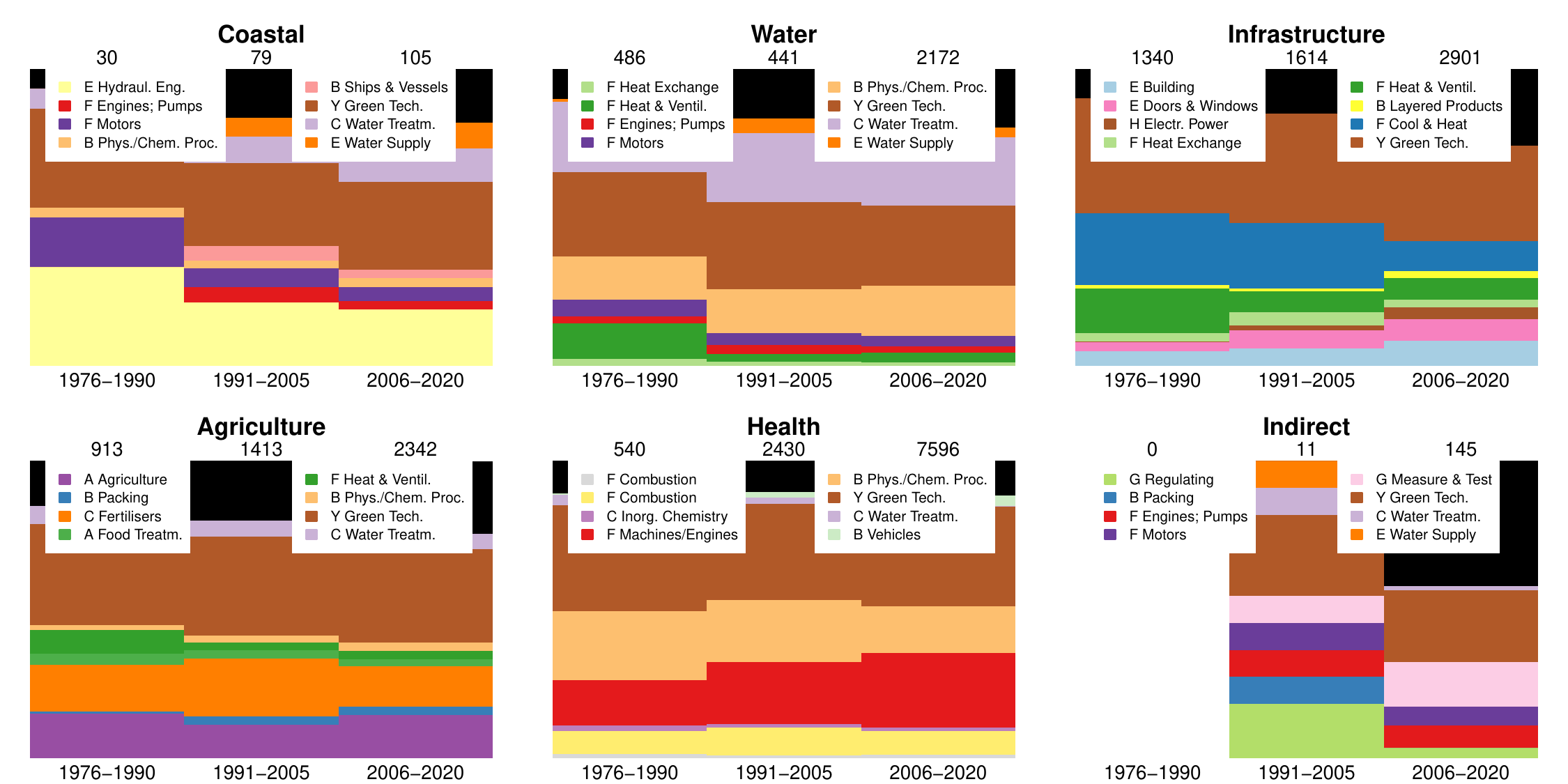}
    \caption{Co-classification of adaptation technologies with mitigation co-benefit}
    \label{fig:SI_convergence_MA_complements_3digit}
    
    }

    \footnotesize
    Notes: This figure shows the 8 most frequent co-classifications (3-digit CPC classes) of adaptation technologies that are simultaneously classified as mitigation technology.  
\end{figure}

\begin{figure}
    {\centering
    \includegraphics[width=\textwidth]{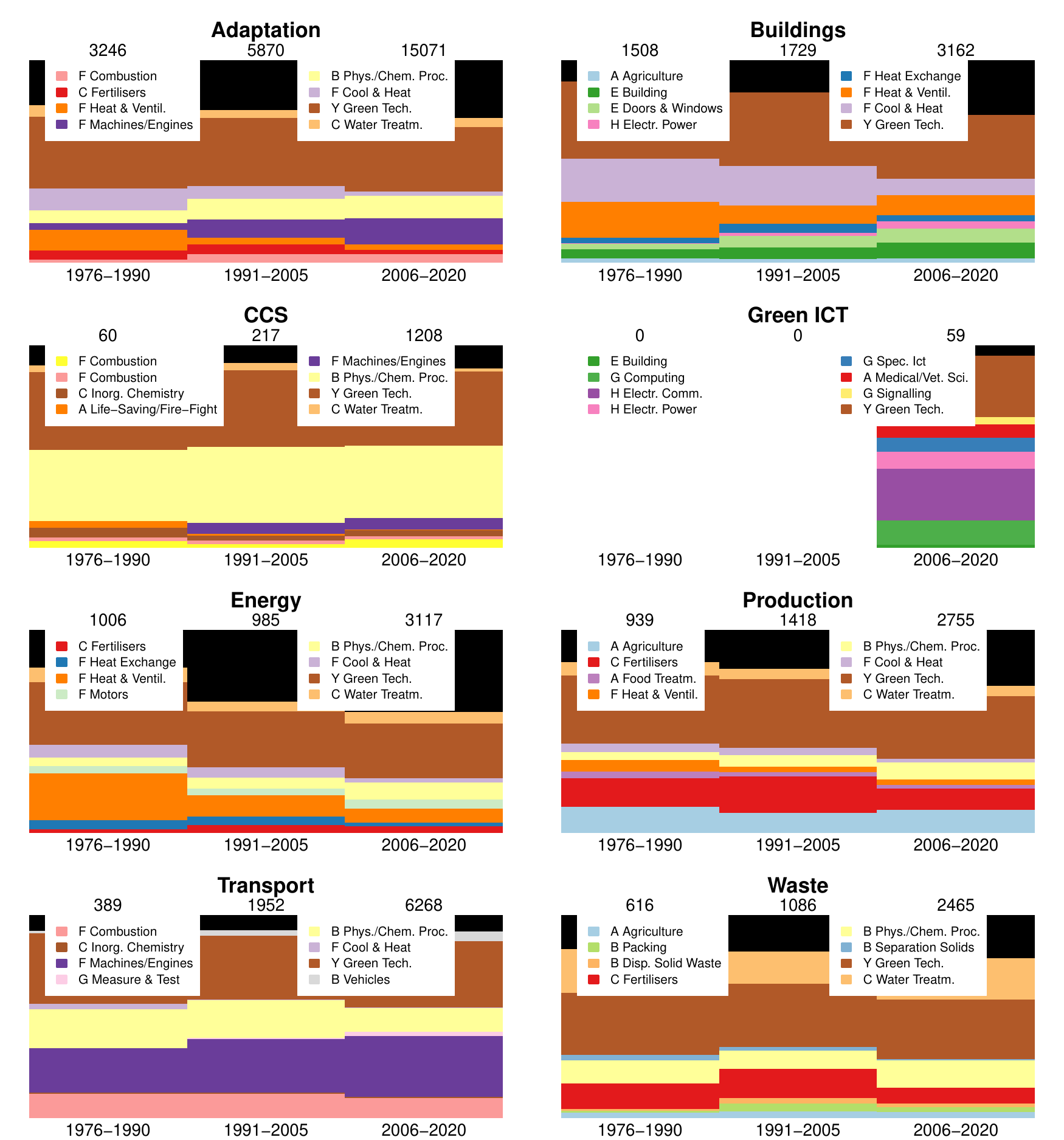}
    \caption{Co-classification of dual-purpose adaptation and mitigation technologies}
    \label{fig:SI_convergence_MA_complements_mitigation_3digit}
    
    }
    
    \footnotesize
    Notes: This figure shows the 8 most frequent co-classifications (3-digit CPC classes) of dual-purpose adaptation and mitigation patents, i.e. patents that are simultaneously classified as adaptation and mitigation technology. 
\end{figure}

\FloatBarrier





\end{document}